\documentclass[twocolumn]{aastex63}
\usepackage{colortbl}
\definecolor{Gray}{gray}{0.9}
\usepackage{multirow}

\shorttitle{Imaging and RV}
\shortauthors{Zhexing Li et al.}

\begin{document}

\definecolor{OG}{rgb}{0,0.6,0}

\title{Direct Imaging of Exoplanets Beyond the Radial Velocity Limit: Application to the HD~134987 System}

\author[0000-0002-4860-7667]{Zhexing Li}
\affiliation{Department of Earth and Planetary Sciences, University of California, Riverside, CA 92521, USA}
\correspondingauthor{Zhexing Li}
\email{zli245@ucr.edu}

\author[0000-0003-0220-0009]{Sergi R. Hildebrandt}
\affiliation{Jet Propulsion Laboratory, California Institute of Technology, Pasadena, CA 91109, USA}
\affiliation{Division of Physics, Mathematics and Astronomy, California Institute of Technology, Pasadena, CA 91125, USA}

\author[0000-0002-7084-0529]{Stephen R. Kane}
\affiliation{Department of Earth and Planetary Sciences, University of California, Riverside, CA 92521, USA}

\author[0000-0001-5484-1516]{Neil T. Zimmerman}
\affiliation{Exoplanets and Stellar Astrophysics Laboratory, NASA Goddard Space Flight Center, Greenbelt, MD 20771, USA}

\author[0000-0001-8627-0404]{Julien H. Girard}
\affiliation{Space Telescope Science Institute, Baltimore, MD 21218, USA}

\author[0000-0002-9032-8530]{Junellie Gonzalez-Quiles}
\affiliation{Department of Earth and Planetary Sciences, Johns Hopkins University, Baltimore, MD 21218, USA}

\author[0000-0002-0569-1643]{Margaret C. Turnbull}
\affiliation{SETI Institute, Mountain View, CA 94043, USA}

%%%%%%%%%%%%%%%%%%%%%%%%%%%%%%%%%%%%%%%%%%%%%%%%%%%%%%%%%%%%%%%%%%%%

\begin{abstract}

Future direct imaging missions will primarily observe planets that have been previously detected, mostly via the radial velocity (RV) technique, to characterize planetary atmospheres. In the meantime, direct imaging may discover new planets within existing planetary systems that have bright enough reflected flux, yet with insufficient signals for other methods to detect. Here, we investigate the parameter space within which planets are unlikely to be detected by RV in the near future due to precision limitations, but could be discovered through reflected light with future direct imaging missions. We use the HD~134987 system as a working example, combine RV and direct imaging detection limit curves in the same parameter space through various assumptions, and insert a fictitious planet into the system while ensuring it lies between the RV and imaging detection limits. Planet validity tested through dynamical simulations and retrieval tests revealed that the planet could indeed be detected by imaging while remaining hidden from RV surveys. Direct imaging retrieval was carried out using starshade simulations for two mission concepts: the \textit{Starshade Rendezvous Probe} that could be coupled with the \textit{Nancy Grace Roman Space Telescope}, and the \textit{Habitable Exoplanet Observatory}. This method is applicable to any other systems and high contrast direct imaging instruments, and could help inform future imaging observations and data analysis on the discovery of new exoplanets.

\end{abstract}

\keywords{planetary systems -- techniques: radial velocity, direct imaging}

%%%%%%%%%%%%%%%%%%%%%%%%%%%%%%%%%%%%%%%%%%%%%%%%%%%%%%%%%%%%%%%

\section{Introduction}
\label{intro}

The radial velocity (RV) technique is one of the most important exoplanet detection methods that led to the first exoplanet discoveries \citep{Latham1989,Mayor1995,Butler1996}. Besides making contributions to exoplanet detections, in the recent decade RV also serves as an important tool to characterize the Keplerian orbital elements as well as to determine the masses of planets discovered by transit techniques, such as those by the Kepler \citep{Borucki2010} and the Transiting Exoplanet Survey Satellite (TESS) missions \citep{Ricker2015}. Discoveries by the Kepler mission \citep{Borucki2011a,Borucki2011b,Batalha2013,Burke2014,Rowe2015,Mullally2015,Coughlin2016,Thompson2018} as well as exoplanet occurrence rate studies \citep{Howard2012,Petigura2013} indicate an increase in population for terrestrial planets, including those within the habitable zone (HZ) \citep{Kane2016}. These planets require mass measurements from RV when transit timing variations are not available for proper characterization of physical properties and interpretation of atmospheric data \citep{Batalha2019}. Although some short period low mass planets could be recovered by RV with careful treatment, those with longer orbital periods would induce RV variations well below the sensitivities of current common spectrographs of a few meters per second \citep{Fischer2016}. With the advent of newly designed spectrographs such as NEID \citep{Schwab2016} and ESPRESSO \citep{Pepe2020} pushing the RV detection limit down to tens of centimeter per second regime, low mass terrestrial planet detections around nearby bright stars become even more tangible thanks to these extreme precision RV instruments. However, without proper treatment of intrinsic stellar activity, noise such as stellar rotation, magnetic cycle, and jitter due to chromospheric activity may still present a challenge to the detection of low mass planetary signals around the detection limits \citep{Fischer2016,Luhn2020,Meunier2019,Meunier2020}. In addition, planets such as those that are further out from the host stars and Earth-sized terrestrial counterparts that lie in the HZ around solar-like stars could still remain below the detection threshold of RV in the near future.

Within the last decade, direct imaging has been evolving at a lightning pace in instrumentation and imaging processing techniques. This method's ability to directly image exoplanets with light reflecting straight from the planet's surface or atmosphere not only makes direct detection of the planet possible, but also allows direct atmospheric retrieval and characterization \citep{Feng2018, Damiano2020}. High contrast direct imaging from both ground and space has successfully discovered several long orbital period giant planets \citep{Chauvin2004,Marois2008,Lagrange2009} as well as measured emission spectra of several exoplanets \citep{Janson2010,Konopacky2013}. However, current application of direct imaging are only limited to young, self-luminous giant planets that emit strong infrared emission, or super-Jupiters that have large angular separation from their host stars and are much inflated in their sizes. These detections typically have contrast ratios around $10^{-4}$ to $10^{-5}$ while a Jupiter and an Earth analog would have contrast ratios on the order of $10^{-9}$ and $10^{-10}$, respectively. Future space mission such as the \textit{Nancy Grace Roman Space Telescope} (\textit{Roman} hereafter, formerly known as the \textit{Wide Field Infrared Survey Telescope}, \citet{Spergel2015}), a 2.4-m telescope, would be able to reach a predicted contrast ratio of a few times $10^{-9}$ to image and characterize the atmospheres of giant planets, thanks to the state-of-the-art starlight suppression enabled by a low order wavefront sensing and control system for the internal coronagraph instrument (CGI) \citep{Trauger2016}. If paired with the \textit{Starshade Rendezvous Probe} (SRP),
the instrument contrast could reach the required 1$\times$10$^{-10}$ performance, or possibly even around $4.0\times10^{-11}$ \citep{Seager2019, Romero-Wolf2021}, making direct imaging of terrestrial exoplanets possible.
More precisely, the instrument contrast of a starshade-telescope instrument 
at some angular distance from the starshade's center
is defined as the amount of light at that angular distance averaged over a resolution element 
(e.g. within a PSF FWHM), divided by the peak 
brightness of the light source as measured by that telescope when there is no starshade in place, 
see e.g.~\cite{Harness2021}. A common choice for the angular distance used
to report the starshade instrument constrast is the starshade (geometric) inner 
working angle (IWA), 
which is the angular radius of the starshade as seen from the telescope --72 milli-arcsec, mas, 
in the blue band direct imaging channel of the SRP that we will consider in this work. 
This is the angular distance that corresponds to the
1$\times$10$^{-10}$ and $4.0\times10^{-11}$ values just mentioned. The \textit{Habitable Exoplanet Observatory} (HabEx, \citet{Gaudi2020}) direct imaging concept, on the other hand, would be a 4-m class space telescope with an external starshade. Although the designed contrast ratio limit at the IWA of 70 mas would be similar to that of SRP, the larger primary mirror aperture, larger optical throughput, broader bandwidth, and greater resolving power could make detecting fainter and smaller terrestrial planets more feasible with HabEx.

Previous studies, including the design reports for both missions, have discussed the occurrence rate of different types of exoplanets and predicted yield of detecting these planets throughout the missions. For the SRP with \textit{Roman}, around 1 to 2 for both Earth and Neptune type planets, and around 2 to 3 for Jupiter-like planets are expected to be detected \citep{Seager2019,Romero-Wolf2021}. The low predicted yield is due to the low optical throughput of the instrument, together with thrust and pointing constraints due to solar avoidance angles, that result in a small number of nearby stars, $\sim20$, to be surveyed by the mission \citep{Feng2018,Seager2019,Romero-Wolf2021}. The predicted yield for HabEx was much higher. In total, 55 rocky planets, 60 sub-Neptunes, and 63 giant planets are expected to be discovered by the HabEx mission \citep{Gaudi2020}\footnote{For details on the assumptions made and how predicted yields were calculated for both mission concepts, please refer to \citet{Seager2019}, \citet{Romero-Wolf2021} for SRP, and \citet{Gaudi2020} for HabEx.}.

The expected mission yields shed light on possible direct imaging discoveries in systems that have known exoplanets previously discovered by other methods. Planets in the target list for characterization by future imaging missions will be mostly RV planets since they have been extensively observed by ground based RV facilities and their Keplerian orbital parameters are well known. However, smaller RV signals caused by potential smaller planets in the system, if there are any, could be hidden below the detection threshold of current RV instruments. In this paper, we used HD 134987 system, one of the more distant future imaging mission target systems (26.2 pc), as a test case to determine if future direct imaging missions could detect low mass planets that are hidden below the RV spectrograph sensitivities as well as to test and compare the imaging capabilities of both \textit{Roman} with SRP and HabEx on detecting such low mass planets at this distance. We did this by injecting a fictitious planet into the system, for which the parameters are determined by combining both RV and direct imaging detection thresholds, carrying out dynamical simulations to determine the likely parameter space within which the fictitious planet could reside, and performing retrieval tests with realistic starshade imaging simulations to confirm it is indeed below RV's detection threshold while it is still detectable by imaging. The use of both RV and direct imaging detection sensitivities in combination with dynamical analysis allows a full exploration of the parameter space within the HD 134987 system for the search of additional potential low mass planetary objects. Although this paper focuses on HD 134987 system only, similar work could be done for other systems and the results could provide valuable information for future imaging missions regarding the chance of detecting an additional new planetary companion within known planetary systems.

In section \ref{limit}, we describe how the detection limits for both RV and direct imaging were calculated and the assumptions considered. In section \ref{stability}, we describe the dynamical simulation performed and the parameters selected for the fictitious planet. In section~\ref{test}, we present the derived expected planet-star flux ratios based on orbital parameters, realistic starshade images and retrieval tests that include both RV and direct imaging. We discuss the results in section \ref{discuss} and conclude in section \ref{conclude}.

%%%%%%%%%%%%%%%%%%%%%%%%%%%%%%%%%%%%%%%%%%%%%%%%%%%%%%%%%%%%%%%%%%%%

\section{Detection Limits}
\label{limit}
A better sense of the relationship between detection thresholds of RV and direct imaging needs to be established before locating the possible regions where additional undetected planets could reside in a system. With increasing synergy between RV and direct imaging field, having one detection limit plot that encompasses detection threshold curves of both methods would be extremely useful and intuitive for realizing the detection sensitivity of known planets and predicting new discoveries. RV detection threshold curves showing the different semi-amplitude sensitivities of the spectrographs are typically displayed in the planetary mass versus orbital period parameter space while for direct imaging, the detection limit curves for different imaging instruments are commonly shown in the planet-star contrast ratio versus angular separation/semi-major axis parameter space. To better visualize and directly compare the detection limits of the two, we made a few assumptions and put both detection limits in planetary radius versus separation parameter space.

    \subsection{Radial Velocity}
    \label{RV}
    For RV detection limit, we used the instrumental precision of Keck HIRES \citep{Vogt1994} and WYIN NEID \citep{Schwab2016} as the two RV standards, where each has a semi-amplitude limit about 1 m/s and 0.3 m/s, respectively. The RV semi-amplitude can be expressed as:
    
    \begin{equation}
        \label{eqn:RV}
        K = \left(\frac{2\pi G}{P}\right)^{1/3}\frac{M_{p}\sin i}{M_{t}^{2/3}}\frac{1}{\sqrt{1-e^{2}}}
    \end{equation}
    where $P$ is the period of the planet, $i$ is the inclination of the orbital plane relative to our line of sight, where an edge-on orbital plane is equivalent to $i$ = 90$^{\circ}$. $e$ is the eccentricity of the planetary orbit and $M_{t} = M_{\ast}+M_{p}$ is the combined mass of the star and the planet.
    
\begin{figure}[tbp]
    %\epsscale{1.29}
    %\plotone{rv.png}
    \includegraphics[trim=9 7 5 3,clip,width=\columnwidth]{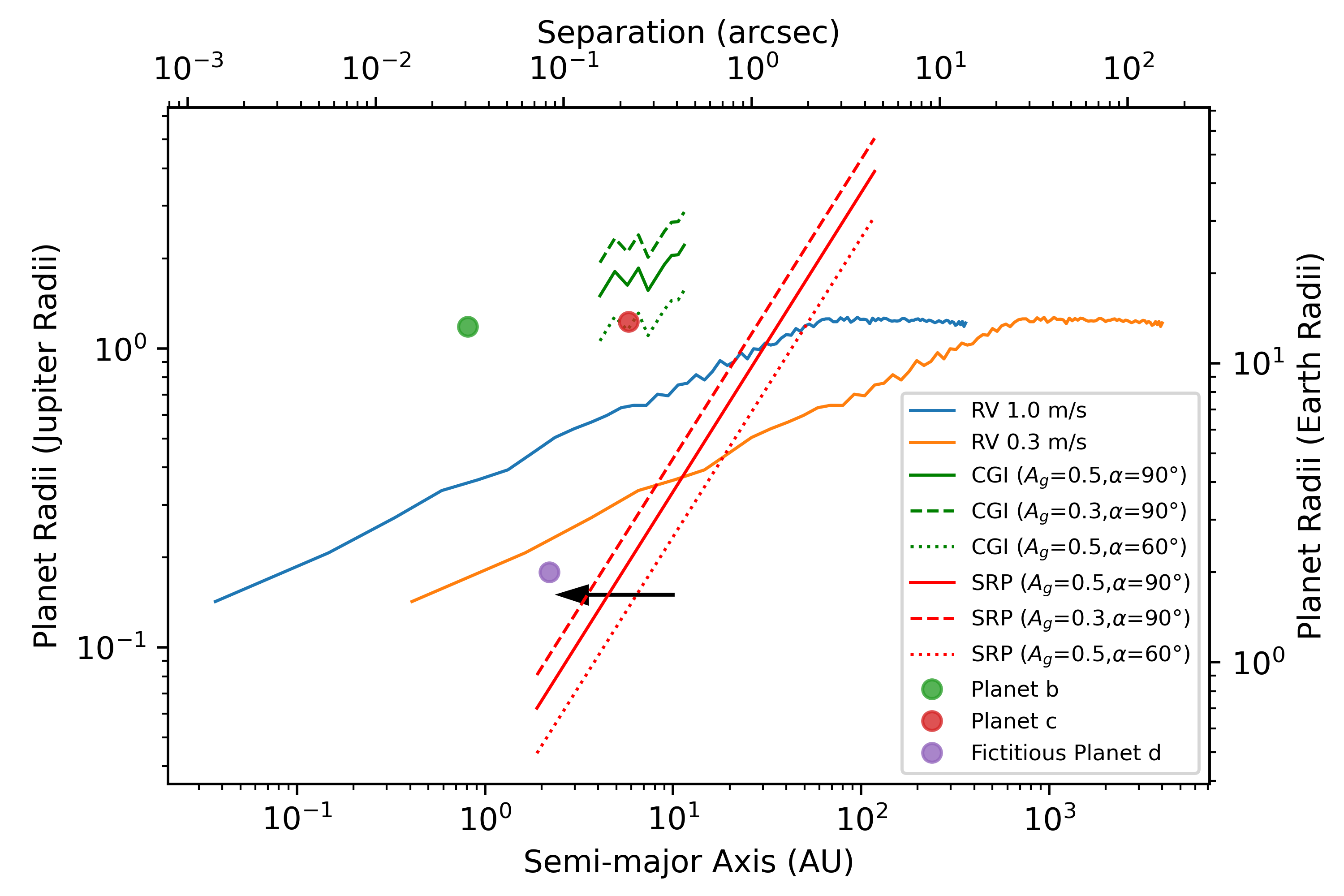}
    \caption{RV and direct imaging detection thresholds of HD 134987 for different instruments in planetary radii and separation space. RV limits are for 1 and 0.3 m/s precisions. Direct imaging thresholds are for \textit{Roman}'s CGI as well as SRP. Shown also are locations of two known planets in the HD 134987 system along with the injected fictitious planet.} 
    \label{fig:threshold}
    \end{figure}

    We constructed the RV detection limit for our test system HD 134987, starting in the planetary mass versus orbital period space. Rearranging equation (\ref{eqn:RV}) for orbital period, we calculated the range of orbital period given $M_p$'s range from 0.01 to 1 Jupiter mass, with a step of 0.01 Jupiter mass. Eccentricity is taken to be zero, and $M_{\ast}$ = 1.10 $M_\odot$, HD 134987's host star mass \citep{Stassun2017}. Varying eccentricity values show insignificant change in the shapes or positions of the RV detection limit curves. Since the inclination for HD 134987 is unknown, we assumed a 45$^{\circ}$ orbital inclination. Orbital period range obtained here can be converted to the corresponding semi-major axis range with ease. With the distance to the system of 26.20 pc \citep{Gaia2018}, angular separation values were also computed. For the range of planetary mass values, we used {\sc Forecaster} \citep{Chen2017}, a forecasting model that is based on a probabilistic mass-radius relation, to convert masses to planetary radii. The result of RV detection limits of 1 and 0.3 m/s in planetary radii and separation space can be seen in Figure \ref{fig:threshold}. Although the RV curves extend into the extremely large separation ($>$ 100 AU), detection in this regime is extremely challenging and the full range of RV curves derived from the input mass range were shown here only for easy comparison between the two sensitivity limits, as well as against the imaging limits described in section \ref{DI}. Higher mass planets would have to orbit further away from the star in order to maintain the same semi-amplitude level (1 and 0.3 m/s in this case) and the RV sensitivity curves flatten out at longer separations. This is due to planets in the Jovian regime with the dominating H-He composition of the envelope are massive enough where gravitational self-compression kicks in to stop the radius from increasing further \citep{Guillot2005}. This is also reflected in Figure 3 of \citet{Chen2017}.
    
    \subsection{Direct Imaging}
    \label{DI}
    We used publicly available performance limit for both \textit{Roman}'s CGI and SRP as the direct imaging limits. CGI detection limit curve is based on the instrumental direct imaging predicted performance at 575 nm with total integration time of 100 hours\footnote{Data files regarding the performances of different direct imaging instruments, including \textit{Roman}'s, are created and hosted by Vanessa Bailey under Github repository: \url{https://github.com/nasavbailey/DI-flux-ratio-plot}.}. For SRP, we used the predicted performance values given in the science report \citep{Seager2019, Romero-Wolf2021}. Both contrast ratio limits are in flux ratio versus angular separation space. The conversion to estimated planetary radii versus separation space employs a Lambert sphere assumption for a planet, which assumes isotropic atmospheric scattering over 2$\pi$ steradians. We used equation \ref{eqn:flux} to estimate the planetary radii $R_{p}$ values based on the provided flux ratio $\epsilon (\alpha ,\lambda )$ values, with the assumptions that the geometric albedo $A_{g}(\lambda )$ takes the values of 0.5 and 0.3, representative of those of Jupiter and Neptune/terrestrial planet analogs in the visible band \citep{Cahoy2010}. Phase function $g(\alpha ,\lambda )$, defined in equation \ref{eqn:phasefn}, was assumed to have a value equivalent to when a planet is at quadrature (phase angle $\alpha$ = 90$^{\circ}$) with respect to the star, at which the planet's location is feasible for carrying out direct imaging observations. This assumption is the same as that in the \textit{Roman} CGI performance predictions. Thresholds for the $\alpha$ = 60$^{\circ}$ and albedo of 0.5 case are also shown for both CGI and SRP to demonstrate the effect of phase angle on planet detectability.
    
    \begin{equation}
        \label{eqn:flux}
        \epsilon (\alpha ,\lambda ) \equiv \frac{f_{p}(\alpha ,\lambda )}{f_{\ast }(\lambda )} = A_{g}(\lambda )g(\alpha ,\lambda )\frac{R_{p}^{2}}{r^{2}}
    \end{equation}
    
    \begin{equation}
        \label{eqn:phasefn}
        g(\alpha ,\lambda ) = \frac{\sin \alpha + (\pi - \alpha ) \cos \alpha }{\pi }
    \end{equation}
    
    The combined detection limit curves of both RV and direct imaging in planetary radii and separation parameter space can be seen in Figure \ref{fig:threshold}. Planets are detectable if they sit above the detection limit curve of a particular method or instrument, or otherwise the retrieval of the planetary signal would be challenging as noise are likely to dominate. The locations of the two known giant planets in the HD 134987 system reveal valuable information: planet c, one of the prime direct imaging targets for future space missions, sits within the field of view (FOV) of both CGI and SRP while planet b orbits well outside the FOV of both and cannot be detected with these future missions even though it reflects bright enough starlight. It is critical to point out from Figure \ref{fig:threshold} that CGI may not be able to detect planet c when the planet is at phase angle of 90$^{\circ}$ due to lower planet-to-star flux ratio at this position. Only at smaller angular separation ($<$60$^{\circ}$ phase angle for example) will planet c become detectable for CGI. SRP however will be able to see the planet throughout its entire orbit. This new way of displaying detection thresholds of RV and direct imaging for one system allows a direct and easy visual comparison of the sensitivities of instruments of different methods and the detectabilities of planets to each. It also gives useful insights to the possible locations where additional planets could be discovered by one method, but not the other. We are interested in finding out if an additional planet may exist in our test system HD 134987 that is below the RV detection threshold, but above the imaging contrast limit. This parameter space of interest in Figure \ref{fig:threshold} lies in the region below the RV 0.3 m/s curve and above the starshade limit (indicated by the arrow), in between the two existing giant planets (hereafter, ``The Region"). Although Figure \ref{fig:threshold} is informative in indicating the presence of such space, the curves were derived from the instrumental limits under the most ideal situations. For a more careful analysis of the allowed parameter space that additional planet may reside in ``The Region", further simulations are needed that take into account dynamical interactions between planets, stellar jitter, spectrograph measurement error, imperfect starshade, local and exozodical light, and detector noise properties. These additional simulations are presented below.

%%%%%%%%%%%%%%%%%%%%%%%%%%%%%%%%%%%%%%%%%%%%%%%%%%%%%%%%%%%%%%%%%%%%

\section{System Stability}
\label{stability}

    \subsection{Dynamical Simulation}
    \label{simulation}

    To explore the dynamical viability of the fictitious planet's orbit in ``The Region", we used {\sc REBOUND} \citep{Rein2012} package to conduct dynamical simulations to search for locations where this fictitious planet may exist. We tested the viable orbits of this planet in between the orbits of the existing two planets using the {\sc MEGNO} (Mean Exponential Growth of Nearby Orbits) indicator within {\sc REBOUND} with the symplectic integrator WHFast \citep{Rein2015}. The {\sc MEGNO} was originally developed by \citep{Cincotta2000} and can be applied to dynamical analysis from galactic scale to planetary systems. For our purpose, the {\sc MEGNO} chaos indicator is useful in distinguishing the quasi-periodic or chaotic orbital time evolution of planetary systems of interest \citep{Hinse2010}. Because we are interested in locating the dynamically stable regions in between the discovered planets, the final {\sc MEGNO} values returned for a grid of provided parameter space after numerically evaluating the {\sc MEGNO} indicator with a given integration time would be useful in determining possible locations where the fictitious planet could dynamically exist. A high {\sc MEGNO} value indicates the chaotic nature of the system and it is unlikely for the orbits to remain dynamically stable over long term.
    
\begin{figure}[tbp]
    \includegraphics[trim=20 10 65 27,clip,width=\columnwidth]{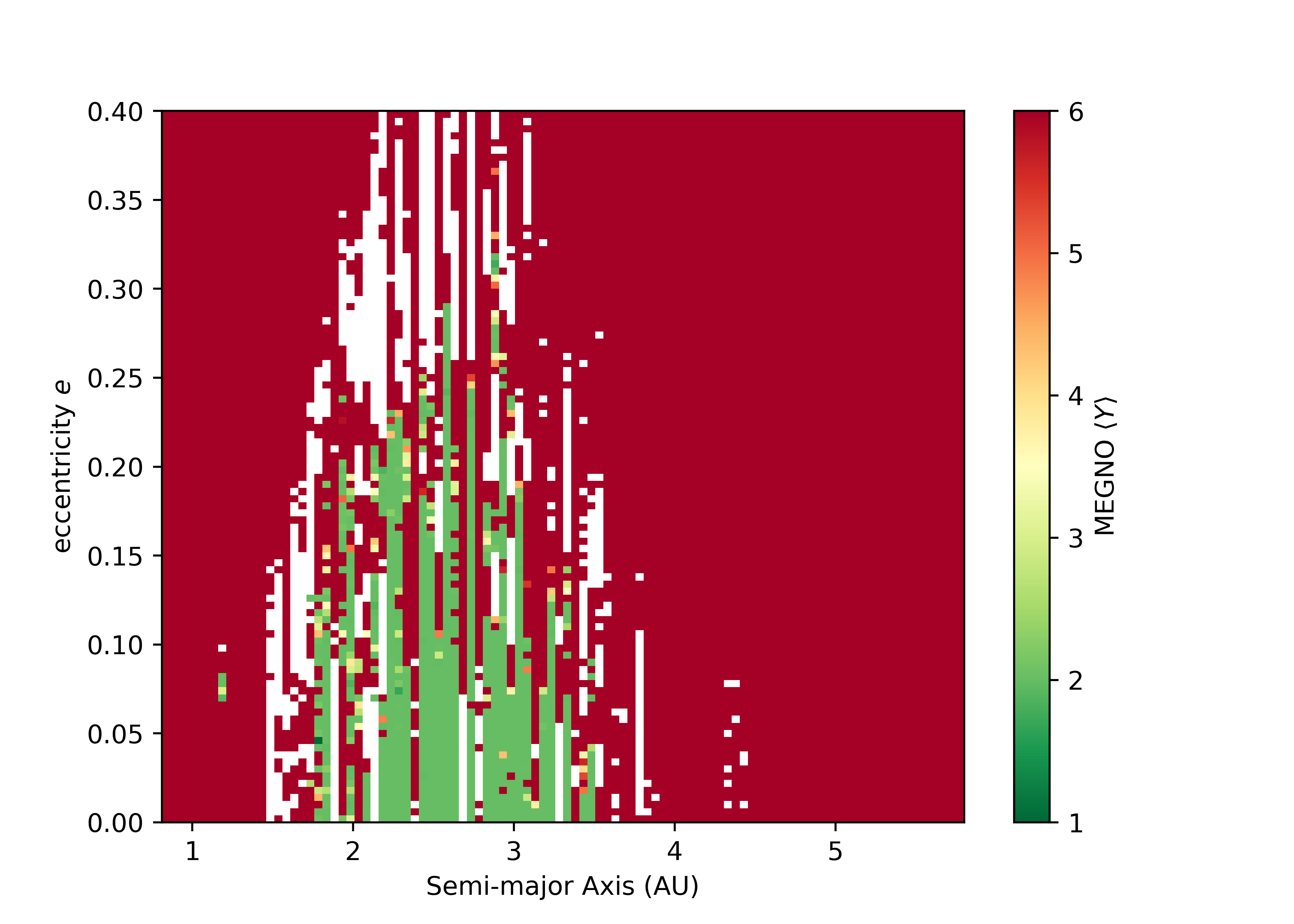}
    \caption{MEGNO simulation result of HD 134987 system over 10 million years with the inclusion of an additional planet in between the existing two. Horizontal and vertical ranges represent the semi-major axis and eccentricity values of the fictitious planet that MEGNO tested. Grids in green indicate non-chaotic result. Red grids mark the locations of high chaos while white ones represents close encounters or collision, both are unfavorable locations for planetary long term stability.} 
    \label{fig:megno}
    \end{figure}

    We selected a radius value of 2 Earth radii from Figure \ref{fig:threshold} for the fictitious planet that fits into the described region, and use the mass-radius relationship tool {\sc Forecaster} \citep{Chen2017} to forecast an approximate mass of 5 Earth masses for this planet, making it a sub-Neptune type of planet. Argument of periastron ($\omega$) for this planet was set to 90$^{\circ}$ and its orbit was assumed to be coplanar with the two known planets. We injected the fictitious planet into the system and ran the MEGNO simulation to test the stochasticity of the system at each grid point of the parameter space we provided. The grid, which can be seen in Figure \ref{fig:megno}, was set up with eccentricity of the additional planet ranging from 0 to 0.4 and semi-major axis values in between the two known planets, from 0.82 to 5.91 AU. Orbital parameters of the two known planets were taken from the published values in \citep{Jones2010}. A summary of the orbital parameters used for the two known planets is in Table \ref{tab:param}. We ran the simulation with an integration time of 10 million years and a time step of 0.03 years (10.95 days), consistent with the recommendation that the time step to be 1/20 of the shortest orbital period in the system \citep{Duncan1998}. The integration was set to stop and return a large {\sc MEGNO} value if the planet was ejected beyond 100 AU. The result of the simulation returns final {\sc MEGNO} values for each specific combination of semi-major axis and eccentricity, with MEGNO value around 2 indicating non-chaotic nature of the system and quasi-periodic motion of the planet if the planet was put at that location \citep{Hinse2010}.

    The result of the {\sc MEGNO} simulation is shown in Figure \ref{fig:megno}. The horizontal and vertical axes are the range of space {\sc MEGNO} simulations have tested. Grid points color coded in green represents {\sc MEGNO} values around 2; red indicates simulations that returned chaotic results; and other events such as closer encounters, collisions that ended up with $NaN$ simulation results are in white. The {\sc MEGNO} result indicates that a 5 Earth mass fictitious planet that we injected into the system is unlikely to render the system chaotic in the long term and the planet could have a stable orbit in between the existing two giant planets. The green grid points shows that the stable orbit of the injected planet could be in between $\sim$1.9 to $\sim$3.5 AU with eccentricities between 0 and $\sim$0.3.

    \begin{deluxetable*}{lllllllll}[tbp]
    \tablecaption{Orbital Parameters of Known and Injected Fictitious Planets in HD 134987.
    \label{tab:param}}
    \tablehead{
        \colhead{Planet} &
        \colhead{$a$ (AU)} &
        \colhead{$P$ (days)} &
        \colhead{$e$} &
        \colhead{$\omega$ (deg)} &
        \colhead{$T_{p}$ (JD)} &
        \colhead{Mass (M$_{J}$)} &
        \colhead{Radius (R$_{J}$)} &
        \colhead{Albedo}
    }
    \startdata
    b  & 0.82 & 258 & 0.23 & 173 & 2460133 & 2.25 & 1.19 & 0.5        \\ \hline
    c  & 5.91 & 5000 & 0.12 & 15 & 2461100 & 1.16 & 1.23 & 0.5         \\ \hline
    d (fictitious) & 2.20 & 1136 & 0.01 & 270 & 2460000 & 
    0.0157 (5 $M_{\oplus}$) & 0.178 (2 $R_{\oplus}$)  & 0.3 \\ \hline
    \enddata
    \tablecomments{Orbital inclinations were assumed to be 45$^{\circ}$ and longitude of the ascending nodes were set to 270$^{\circ}$ for all planets. Stellar mass takes the value of 1.1 M$_{\odot}$ \citep{Stassun2017} and distance to the system is 26.20 pc \citep{Gaia2018}. Parameters for planet b \& c were taken from \citep{Jones2010}, except for radius values, which were estimated by {\sc Forecaster} \citep{Chen2017}. The host star has a spectral type of G6IV-V \citep{Gray2006} and a V band magnitude of 6.46 \citep{Hog2000}.}
    \end{deluxetable*}

    \subsection{Fictitious Planet Location}
    \label{location}
    We selected the semi-major axis and eccentric values for the fictitious planet according to the result of the {\sc MEGNO} simulation and the location of ``The Region" in Figure \ref{fig:threshold}. The dynamically favored semi-major axis range of 1.9 to 3.5 AU according to the {\sc MEGNO} simulation result sits comfortably within the ``The Region" in Figure \ref{fig:threshold} where it is below the RV thresholds but above those of SRP's. Locations in Figure \ref{fig:threshold} near the detection limit curves would be subject to the noise contamination resulting in much longer integration time since they are near the instrumental limits. To avoid noise issues and ensure our injected planet to be detected by imaging as easily as possible, we selected the semi-major axis value for our planet to be 2.2 AU so that it resides well clear of the detection thresholds of SRP. At this distance, eccentricity could take any value up to around 0.2 as Figure \ref{fig:megno} indicated. For simplicity, we assign a near circular eccentricity of 0.01 for the fictitious planet. Other parameters are kept unchanged and a summary of orbital parameters of the fictitious planet are listed in Table \ref{tab:param} and its exact location in Figure \ref{fig:threshold} is plotted as well. We also ran an N-body simulation within {\sc REBOUND} using the exact same parameters in Table \ref{tab:param} to check the time evolution of all orbits and obtained stable results for all planets over 10 million year timescale.

%%%%%%%%%%%%%%%%%%%%%%%%%%%%%%%%%%%%%%%%%%%%%%%%%%%%%%%%%%%%%%%%%%%%

\section{Retrieval Tests}
\label{test}
To further verify the injected fictitious planet within ``The Region" in Figure \ref{fig:threshold} in the HD 134987 system is indeed below the RV and above the direct imaging detection limits, we conducted retrieval tests for each method. For each one, we predicted the chance of detection of the planet for each method based on detection limit equations, and then carried out the retrievals with realistic noise components.

    \subsection{Radial Velocity Retrieval}
    \label{rvtest}
        \subsubsection{Detectability}
        \label{RVdetect}
        The RV semi-amplitude ($K$) of the star's induced oscillation by the planet around it usually is a good indicator of whether the planet could be detected by RV instruments with certain sensitivities that could reach down to a minimum $K$ value. Using the orbital parameters of the fictitious planet in Table \ref{tab:param} and equation \ref{eqn:RV}, we calculated the semi-amplitude induced by this planet to be about 0.203 m/s. This value is below the lowest nominal semi-amplitude limit of 0.3 m/s for current RV instruments that we employ, as indicated in Figure \ref{fig:threshold}, and suggests that the planet would be extremely difficult to be retrieved especially with the addition of noise.
        
        \subsubsection{Retrieval}
        \label{RVretrive}
        The original RV data set published in \cite{Jones2010} contains 138 observations taken by HIRES and UCLES spectrographs on Keck and Anglo-Australian Telescope. The data set spans almost 5000 days with instrumental precisions on the order of 1 and 2 m/s, respectively, and no additional signals were found other than the two known planets. To test whether RVs with higher instrumental precision around 0.3 m/s can recover the signal of our fictitious planet, we created a synthetic RV data set with a time range of 5000 days and a total of 150 data points, similar to the original data set. To mimic the uneven sampling of RV, we increased the number of evenly separated time stamps to 50\% more than the originally intended 150 data points, totalling it to 225. These time stamps were then passed through a Gaussian filter with a sigma of 10 days to increase the variation in between time stamps. The synthetic RV at each time stamp were then calculated for each planet by solving the Kepler's equations using the orbital parameters of all three planets listed in Table \ref{tab:param} and equation \ref{eqn:velocity}, where $V_{0}$ is velocity offset, which is taken to be zero in our case, and $f$ is the true anomaly of the planet, calculated numerically at each time step for each planet:
        
        \begin{equation}
            \label{eqn:velocity}
            V = V_{0} + K[\cos (\omega +f) + e \cos \omega ]
        \end{equation}
        
\begin{figure}[tbp]
        \includegraphics[trim=15 5 40 30,clip,width=\columnwidth]{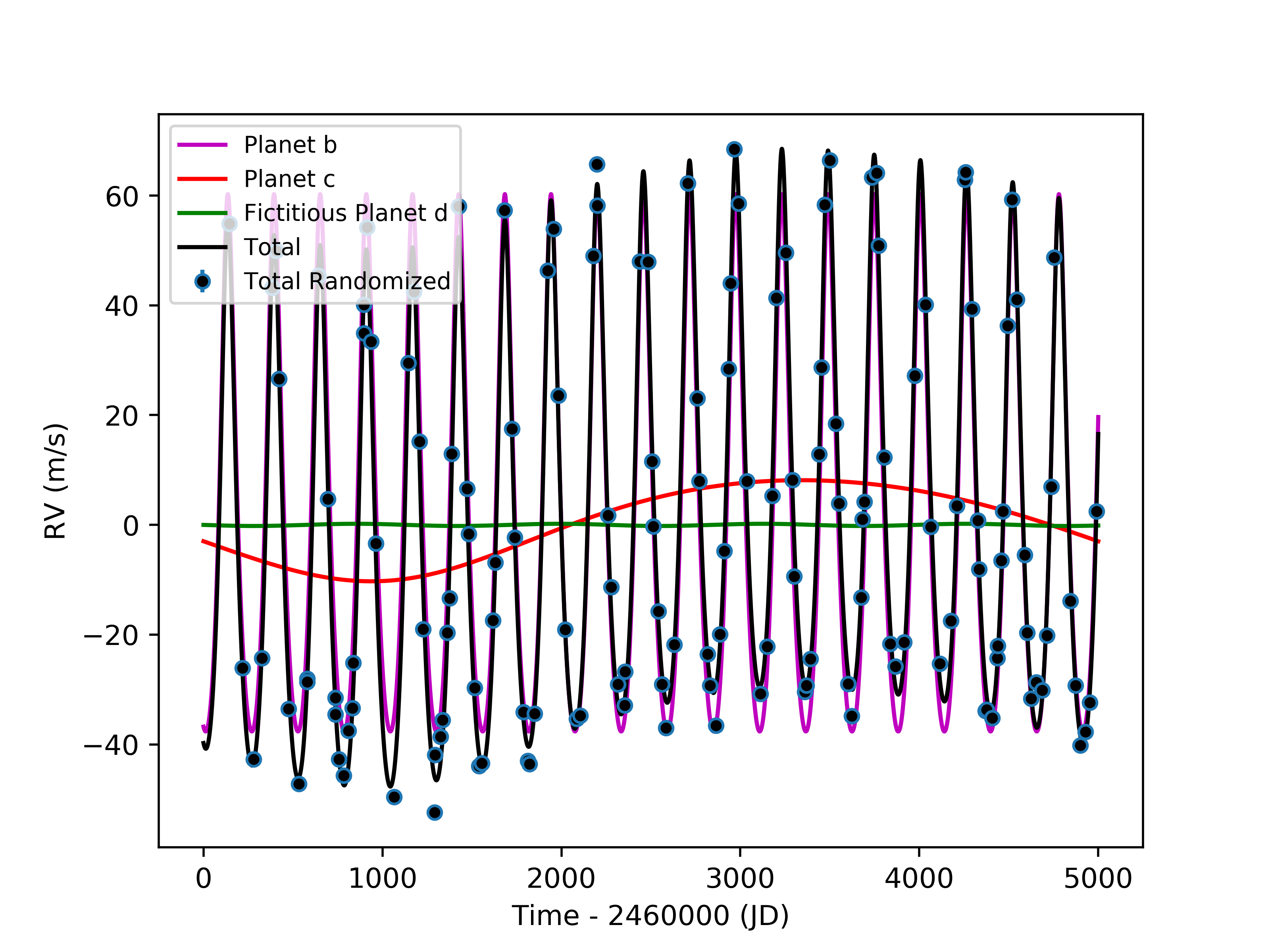}
        \caption{Synthetic RV dataset to be used for retrieval. Data points are randomized to include instrumental and astrophysical noise. Black curve shows the total RV variation and each colored curves represent velocity contributions from the individual planets. Error bars for data points are too small to be seen in the plot.} 
        \label{fig:synrv}
        \end{figure}

        The total RV for the system is simply the superposition of the individual velocity components from each planet. To mimic the RV scatter due to noise, all the data points were passed through another Gaussian filter, with the sigma variation being the quadrature sum from instrumental uncertainty of 0.3 m/s and host star jitter of 3.5 m/s \citep{Butler2006}. Finally, the additional number of 75 RV data points introduced at the beginning were taken out at randomly selected time stamps to further increase the randomness in sampling for the rest 150 RV points. The final synthetic RV data points created as well as the velocity components from each planet are shown in Figure \ref{fig:synrv}.
        
        \begin{figure*}[tbp]
        \centering
        \includegraphics[trim=30 0 30 10,clip,width=0.8\textwidth]{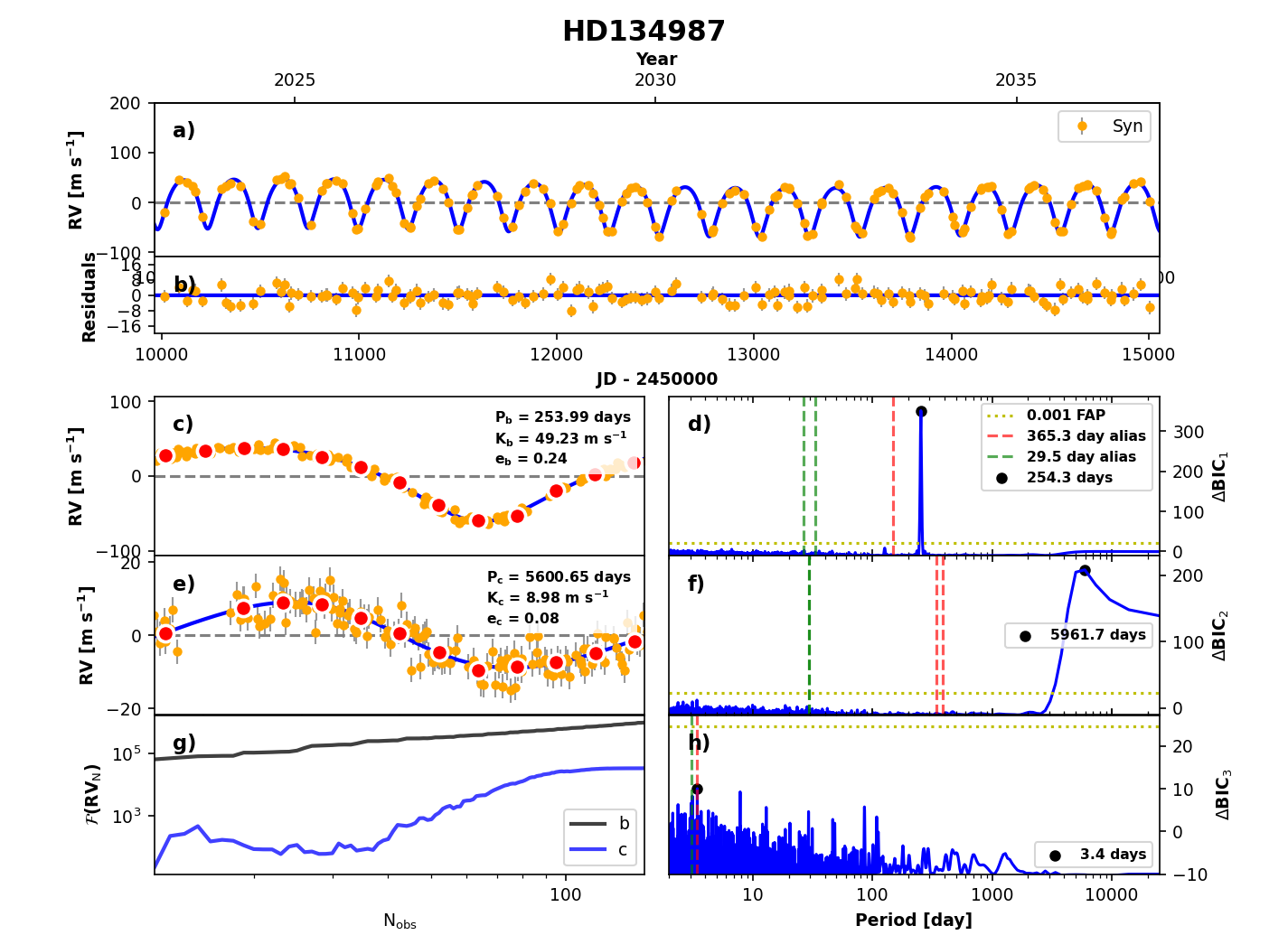}
        \caption{RVSearch result of the synthetic RV dataset. Panel (a) and (b) shows the best fit of the dataset and the residual. Bottom panels show the phase-folded fit for the planets detected by the search on the left and their corresponding periodogram search results. Two known giant planets are clearly recovered by RVSearch. But there is no apparent signal for the injected fictitious planet.} 
        \label{fig:rvsearch}
        \end{figure*}
        
        The RV retrieval of the fictitious planet was carried out first by passing the dataset through a periodogram to search for any significant periodic signals. We used the package RVSearch (Rosenthal et al. submitted), an RV periodogram tool that iteratively search for periodic signals in the time series RV dataset given a grid of period space. It uses the change in the Bayesian Information Criterion ($\Delta$BIC) between the model at the current grid and the best fit model as a measure of the goodness of the fit. The signal was considered significant if it peaks above the 0.1 \% false alarm probability (FAP) level. Figure \ref{fig:rvsearch} shows the result of the search. It is apparent that the two known giant planets were successfully recovered by the search, well above the 0.1\% FAP level. However, RVSearch did not find any significant periodic signals in the residual of the two-planet fit and there is no peaks around the period of the injected fictitious planet.
        
        In addition to the periodogram search, we also used RV modeling toolkit RadVel \citep{Fulton2018} to search for the third signal in the synthetic dataset. RadVel employs maximum a posteriori optimization for RV fitting and Markov Chain Monte Carlo (MCMC) with robust convergence criteria for confidence interval estimation. For the retrieval, we provided orbital parameter priors for all three planets using values from Table \ref{tab:param} and fitted for the three-planet model. The MCMC was run with eight independent ensembles in parallel with 50 walkers per ensemble, where each walker can take up to 10000 steps. The run is considered finished if all the convergence criteria are met or the maximum number of steps for all the walkers are achieved. The MCMC run on the three-planet model finished with maximum number of steps taken and the convergence was not met, indicating the inclusion of the third planet may not be a good fit to the model. With the results from RVSearch and RadVel, we concluded the injected fictitious planet that lies below the RV instrumental precision cannot be successfully retrieved.

    \subsection{Direct Imaging Retrieval}
    \label{DItest}
        \subsubsection{Detectability}
        \label{DIdetect}
        A planet-star flux ratio can be estimated with a Lambert sphere assumption using equation \ref{eqn:flux}, \ref{eqn:phasefn}, and \ref{eqn:alpha} (below), where $\alpha$ is the phase angle of the planet, $\omega$ and $f$ being the argument of periastron and true anomaly of the planet, and $i$ is the inclination of the orbit. 
        
        \begin{equation}
            \label{eqn:alpha}
            \cos \alpha = \sin(\omega + f)\sin i
        \end{equation}
        
        We chose geometric albedo of 0.5 for both known giant planets b \& c and 0.3 for the fictitious planet. The chosen albedo values are consistent with previous estimates and models for Jupiter and Neptune analogs \citep{Cahoy2010}. Although planet b orbits much closer to the host star than a typical cold gas giant and its geometric albedo value is likely to be much lower than 0.5 \citep{Kane2010}, its detection in direct imaging products of either SRP or HabEx is unlikely because its projected orbit lies well outside the starshade FOV for both SRP and HabEx.
        
\begin{figure*}[htbp]
            \begin{center}
                \begin{tabular}{cc}
                    \includegraphics[trim=40 3 10 20,clip,width=0.5\textwidth]{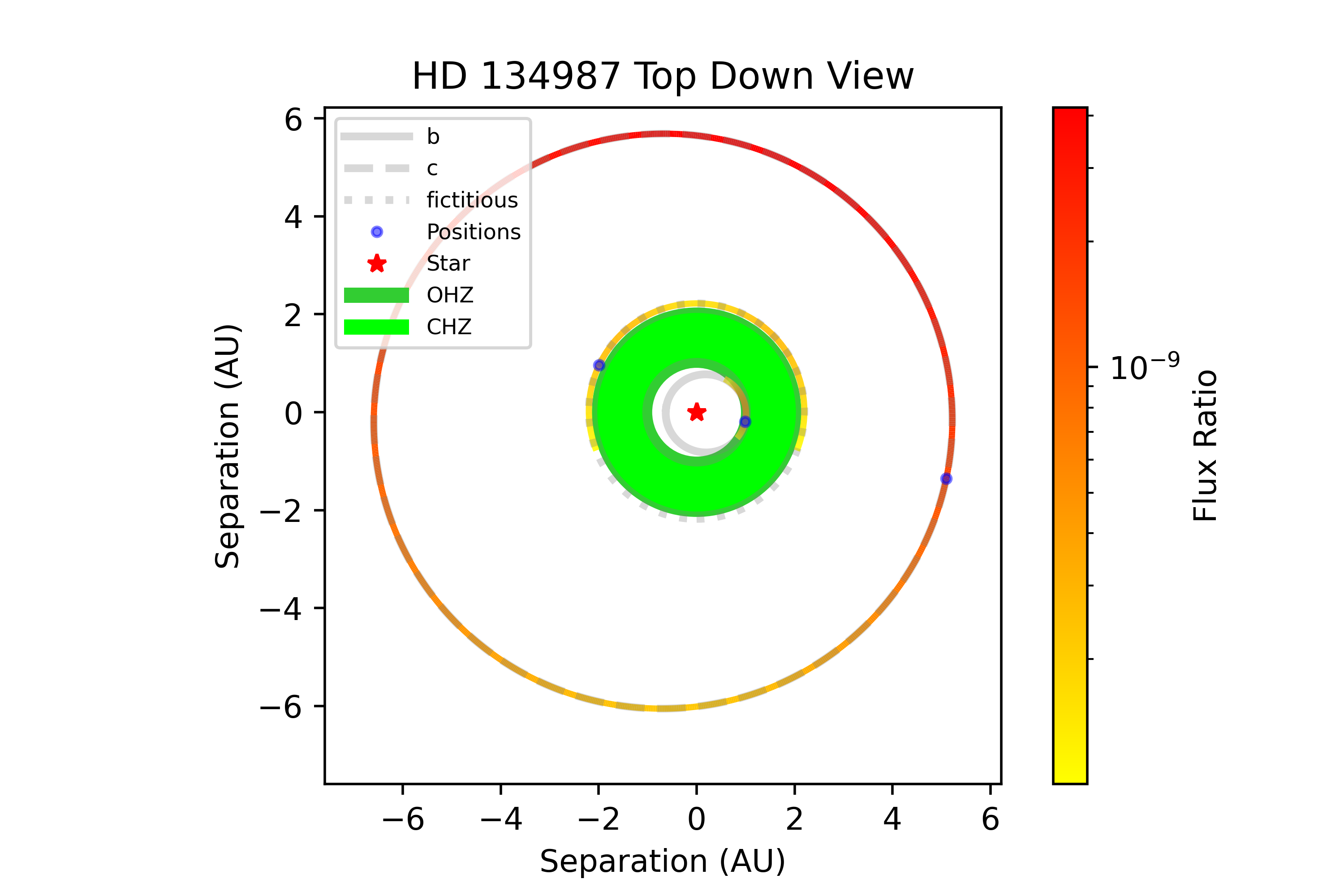} &
                    \includegraphics[trim=40 3 10 20,clip,width=0.5\textwidth]{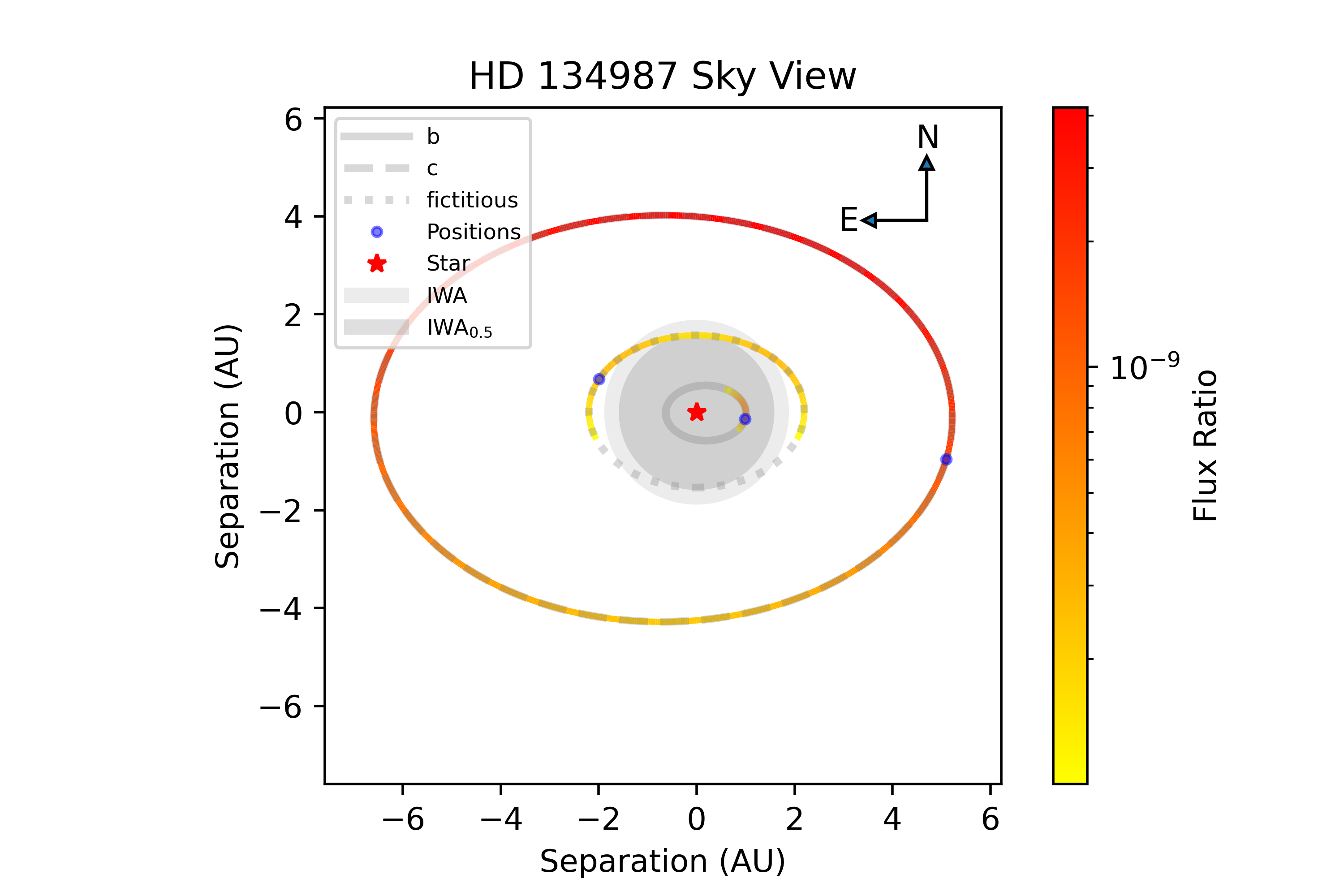} \\
                \end{tabular}
            \end{center}
            \caption{Top down view (left) and sky projection view (right) of the HD 134987 system for \textit{Roman}'s SRP. Planet b \& c are known giant planets and the injected fictitious planet sits in between the two. Orbits are color coded at locations where the calculated flux ratio values are above the 1$\times$10$^{-10}$ flux ratio limit. The sizes of IWA and IWA$_{0.5}$ can be seen in the sky projection view, where the inner b planet is completely covered by IWA$_{0.5}$ while the fictitious planet crosses in and out of the IWA. Tiny dots on all three orbits indicate the location of the three planets when retrievals are carried out.}
            \label{fig:orbit}
        \end{figure*}

\begin{figure}[htbp]
        \includegraphics[trim=20 0 30 20,clip,width=\columnwidth]{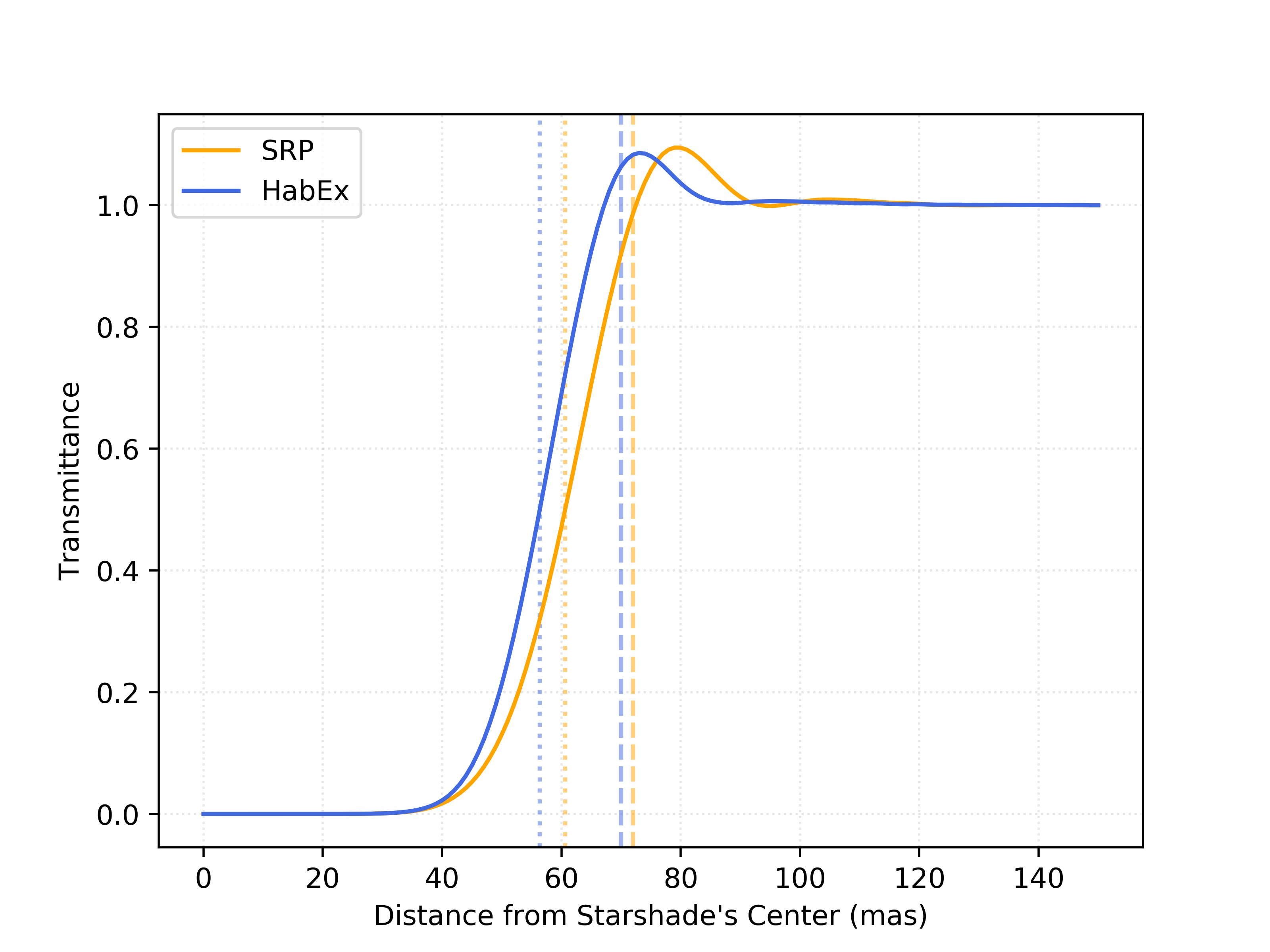}
        \caption{Transmittance profiles of starshades averaged over respective bands for both \textit{Roman}'s SRP and HabEx mission concepts. The vertical dashed and dotted lines are the locations of IWA and IWA$_{0.5}$ for both instruments. Values of the transmittance above 1 near the IWA are a known effect of Fresnel diffraction, which is the diffraction regime that applies to the starshades considered in this work.} 
        \label{fig:tau}
        \end{figure}

\begin{figure}[htbp]
        \includegraphics[trim=17 2 35 15,clip,width=\columnwidth]{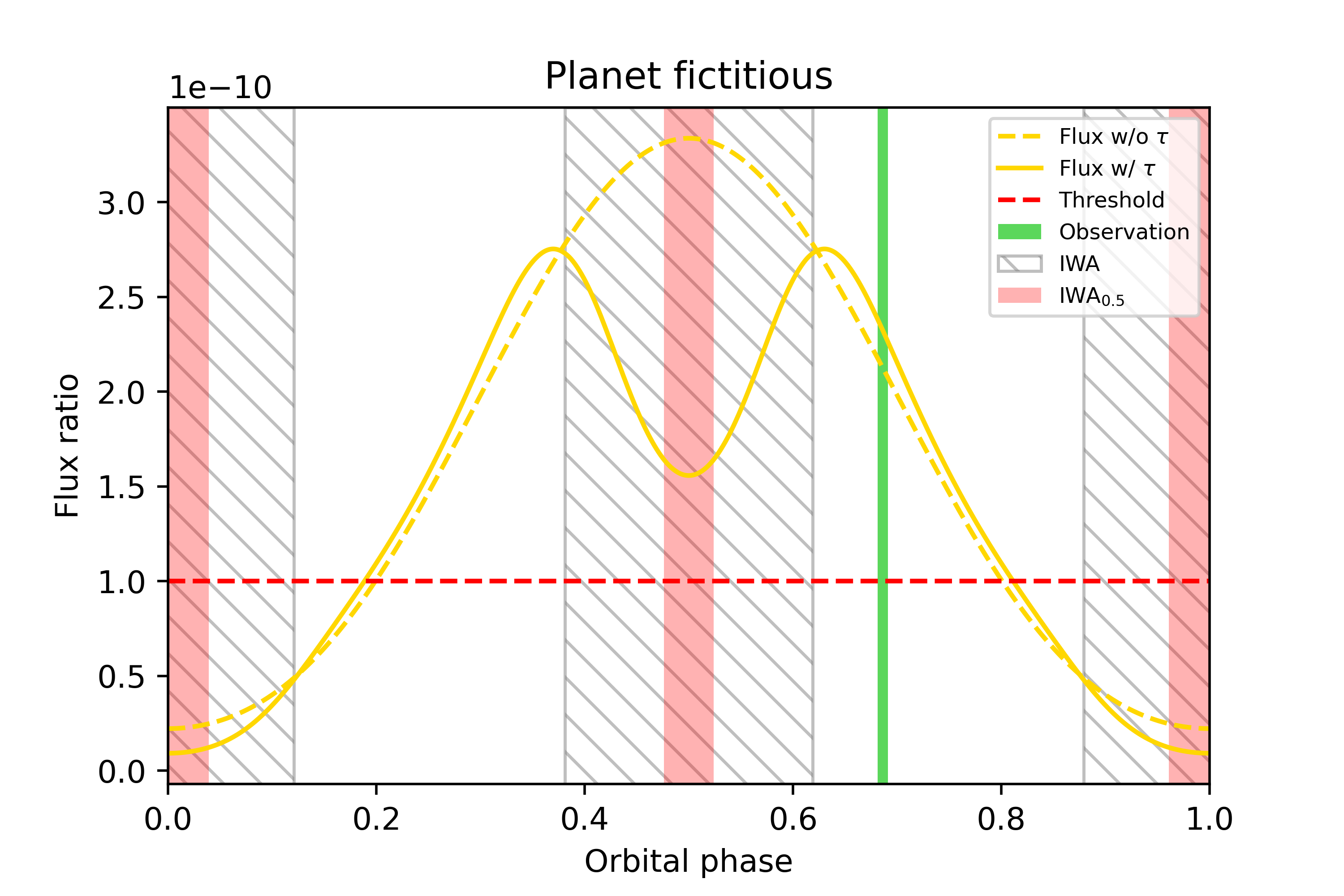}
        \caption{Individual flux variation plot for the fictitious planet in the case of SRP as it makes one complete orbit around the host star. Orbital phase zero here is defined as the passage of periastron. Both fluxes with (solid) and without (dashed) the presence of starshade are displayed to show the difference caused by starshade's transmittance ($\tau$). Phases covered by IWA and IWA$_{0.5}$ are marked with grey hatches and light red band, respectively. The green band indicates the time/phase when the retrievals are carried out.} 
        \label{fig:visibility}
        \end{figure}

        We created a code that visualizes the orbits of planetary systems in both top down and sky projection views, where planetary orbits are color coded with the planet-star flux ratio that are above the provided instrumental contrast limit as seen from Earth at each orbital position based on Lambert sphere flux ratio calculations (Figure \ref{fig:orbit}). The top down view also includes a visualization of conservative and optimistic HZ (CHZ and OHZ, respectively), for which the boundaries are defined in \citet{Kopparapu2013,Kopparapu2014} and the sky view includes the projection of both geometric inner working angle (IWA) and the IWA$_{0.5}$. 
        The definition of the IWA was given in section \ref{intro} and is simply the angular radius
        of the starshade as seen from the telescope. On the other hand, 
        the IWA$_{0.5}$ is defined as the angular radius 
        from the starshade center where the starshade transmittance is 50\%, see Figure~\ref{fig:tau}. That is, where the ratio between the flux of a point-like source with the starshade in place or without it is 50\%. Given a starshade-telescope distance, the IWA does not depend on the wavelength, whereas the IWA$_{0.5}$ does, although in a weak fashion. For instance, across the whole visible passband of HabEx (450-975 nm), its value changes by a 4\%, see e.g. Figure 6.4-3 in~\citet{Gaudi2020}. In the sequel we adopt the average value of IWA$_{0.5}$ across the passband for both SRP, IWA$_{0.5}=60.7$ mas, and HabEx, IWA$_{0.5}=56.4$ mas. Planets that orbit within the IWA$_{0.5}$ and IWA could still be detected. However, below the IWA$_{0.5}$, optical effects of an imperfect starshade that are not adequately simulated could impact the conclusions and it is not considered in the literature. Thus, the IWA$_{0.5}$ could be considered a reasonable proxy for the minimum separation from the star for exoplanet detection \citep{Gaudi2020}. The code also makes an alternative view of showing planets' visibility by plotting out individual planet's flux variation throughout the entire orbital phase, and overplotted with the contrast ratio detection threshold as well as IWA and IWA$_{0.5}$ of the instrument to show the visibility of the planet throughout the orbit (Figure \ref{fig:visibility}). Occulter transmittance ($\tau$) can also be incorporated for more accurate flux ratio calculations. If provided with time epochs for observation, planetary positions can be overplotted on all plots indicating the locations of all planets and estimated flux ratios during the time of observation. A detailed summary including time of observation, orbital phase, phase angle, sky projection coordinates, angular separation, and flux ratio of all planets in the system can also be saved in a separate text file.
        
        Using this code along with orbital parameters in Table \ref{tab:param}, we estimated the detectability of our fictitious planet through \textit{Roman}'s SRP and HabEx imaging missions. We left \textit{Roman}'s CGI out of consideration because the detection of the fictitious planet is far beyond the capability of the instrument (Figure \ref{fig:threshold}). Shown in Figure \ref{fig:orbit} is the orbit and flux ratio plots of HD 134987 system with the fictitious planet for the \textit{Roman}'s SRP mission concept in both top down and sky projection view. Similar plots for HabEx are not shown here because the differences in fluxes from SRP are indistinguishable by eye. For SRP mission, we assumed usage of the blue band (425-552 nm) for planetary detection with an IWA of 72 mas~\citep{Hildebrandt2021}. For HabEx, we used its broad visible channel with 450-975 nm passband with IWA of 70 mas. Instead of using the ideal 4.0$\times$10$^{-11}$ as the minimum flux ratio limit, we used the more conservative 1.0$\times$10$^{-10}$ required minimum detection flux ratio for both SRP and HabEx in anticipation that the imperfection in the manufacturing of the starshade would potentially raise the contrast limit \citep{Seager2019, Gaudi2020,Romero-Wolf2021}. For our direct imaging retrieval, we assumed IWA$_{0.5}$ as the reference limit beyond which one can make planetary detection. To properly account for the flux calculations inside and near the edge of the IWA, we incorporated the transmittance profiles of the starshades for both Rendezvous (425-552 nm) and HabEx (450-975 nm) cases (Figure \ref{fig:tau}).
        
        The inner b planet, as expected, is completely covered by the IWA$_{0.5}$ of the instrument and thus not visible even though for some parts of the orbit the flux ratio is still very high. The outermost planet is bright and separated enough from the host star that the entire orbit is above the threshold and is visible. The fictitious planet orbits at the outer edge of the optimistic HZ and crosses in and out of the IWA due to orbital inclination (Figure \ref{fig:orbit}). The reflected flux of the fictitious planet varies throughout the orbit where it is bright enough for most part of the orbit and falls below the instrumental contrast threshold when near inferior conjunction. The bright solid yellow curve in the individual flux ratio plot (Figure \ref{fig:visibility}) for the fictitious planet reveals that the planet would cross inside the IWA$_{0.5}$ during its conjunction phases of the orbit. Beyond the IWA$_{0.5}$, the planet is still bright enough to be detected despite being inside the IWA. The planet can be seen to achieve its peak apparent brightness at positions slightly beyond the edge of the IWA, reaching an apparent flux ratio value of around 2.8$\times$10$^{-10}$ for SRP. For comparison, planet-to-star flux ratio throughout the entire orbit without the starshade's occultation is shown in Figure \ref{fig:visibility} as bright dashed yellow curve. Flux ratio and planet detectability estimates were also carried out for HabEx but are not shown here due to similar results to SRP, except for configurations of the IWA, IWA$_{0.5}$ and the flux estimates around its edge. The fictitious planet is bright enough for HabEx detection for most of its orbit and again reaches peak brightness of around 3$\times$10$^{-10}$ beyond the edge of IWA.
        
        \begin{deluxetable}{lll}[tbp]
        \tablecaption{Search completeness of the fictitious planet at different flux ratio thresholds.
        \label{tab:percent}}
        \tablehead{
        \colhead{Flux ratio threshold ($\times$10$^{-10}$)} &
        \colhead{\textit{Roman} (\%)} &
        \colhead{HabEx (\%)}
        }
        \startdata
        1.0 & 56 & 61         \\
        1.25 & 50 & 55         \\
        1.5 & 45 & 50         \\
        1.75 & 39 & 45         \\
        2.0 & 31 & 40         \\
        2.25 & 24 & 36         \\
        2.5 & 16 & 25         \\
        2.75 & 2 & 17         \\
        3.0 & 0 & 7      \\ \hline
        \enddata
        \tablecomments{Completeness is indicated by the percentage of the planet's orbit that have flux ratios above the different thresholds being considered. Calculations are for the \textit{Roman}'s SRP and HabEx Starshade observations. Apparent flux ratios lower than 1.0$\times$10$^{-10}$ are not included due to too long integration time needed. Peak apparent flux ratio of the fictitious planet can reach 2.75$\times$10$^{-10}$ for SRP and 3.07$\times$10$^{-10}$ for HabEx.}
        \end{deluxetable}
        
        Both SRP and HabEx may conduct blind search of new exoplanet candidates within known systems. The chance of detecting an undiscovered companion through such search is not only affected by its peak brightness, but the percentage of an orbit the companion could stay above a certain flux ratio threshold and be visible to the instrument. To determine the direct imaging search completeness for the fictitious planet, we injected the planet at different locations in the orbit and calculated the expected flux ratio values at those locations. The planet is considered detected if its flux ratio at the injected location is higher than certain flux ratio thresholds. We computed this for 9 flux ratios thresholds from 1.0$\times$10$^{-10}$ to 3.0$\times$10$^{-10}$. Table \ref{tab:percent} lists the percentage of fictitious planet's orbit where its flux ratio is higher than the different thresholds considered for both SRP and HabEx starshade configurations. A higher percentage indicates the fictitious planet would be more likely to be detected through an imaging search for the system. Although flux ratios below 1.0$\times$10$^{-10}$ could be detected by both instruments in optimistic scenarios, the required integration time becomes closer to the available observing window due to solar avoidance angles for SRP \citep{Seager2019, Romero-Wolf2021}. As can be seen in Figure \ref{fig:orbit}, \ref{fig:visibility}, and Table \ref{tab:percent}, the fictitious planet is clearly bright enough to be above the nominal instrumental detection thresholds for both SRP and HabEx. As expected, the fictitious planet at its brightest times only cover a tiny fraction of its orbit, making detection at its peak brightness unlikely. However, the chance of detection increases as the flux ratio thresholds decreases, for which at least half of the planet's orbit can be detected when the planet has a flux ratio of at least 1.25$\times$10$^{-10}$ for \textit{Roman}, and 1.5$\times$10$^{-10}$ for HabEx. It is worth noting that at each flux ratio threshold, the planet consistently has a large chance of detection for HabEx observations than for SRP, thanks to its slightly smaller IWA and different transmittance profile than SRP's (Figure \ref{fig:tau}).
        
        \subsubsection{Retrieval}
        \label{DIretrieve}
        
        A more realistic detection estimation should take into account different noise sources from both astrophysical and instrumental such as background stars and sky noise, residual starlight, detector's noise, clock induced charge, quantum efficiency, optical throughput, etc. For direct imaging retrieval of the fictitious planet, we carried out simulations using the SISTER~\citep{Hildebrandt2021} package to create and retrieve the planet's signal with both SRP and HabEx configurations. SISTER is a versatile toolkit that provides accurate models of exoplanet images when observed with a starshade. The tool allows inputs of various observational parameters including both astrophysical and instrumental for realistic starshade simulations. It has been used to generate the SRP and HabEx simulations of the Starshade Exoplanet Data Challenge~\citep{Hu2021}.
        
        For the SISTER simulations, we selected the times when the fictitious planet has apparent flux ratios around 2.25$\times$10$^{-10}$. At this brightness, the integration time of both instruments are well below the maximum observing times while ensuring there is still a good chance of detecting the fictitious planet through a blind search within the system. According to Table \ref{tab:percent}, the planet would have a brightness above this level for about a quarter of its orbit. We used this same flux ratio for both SRP and HabEx for direct side-by-side comparison of the imaging capabilities of both instruments. We further calculated the positions of the fictitious planet when this flux ratio would occur to make sure the planet is not within the IWA$_{0.5}$. The positions can be seen as blue dots in Figure \ref{fig:orbit} and vertical green band in Figure \ref{fig:visibility}.
        
\begin{figure*}[htb!]
            \begin{center}
                \begin{tabular}{cc}
                    \includegraphics[trim=20 0 1 40,clip,width=0.48\textwidth]{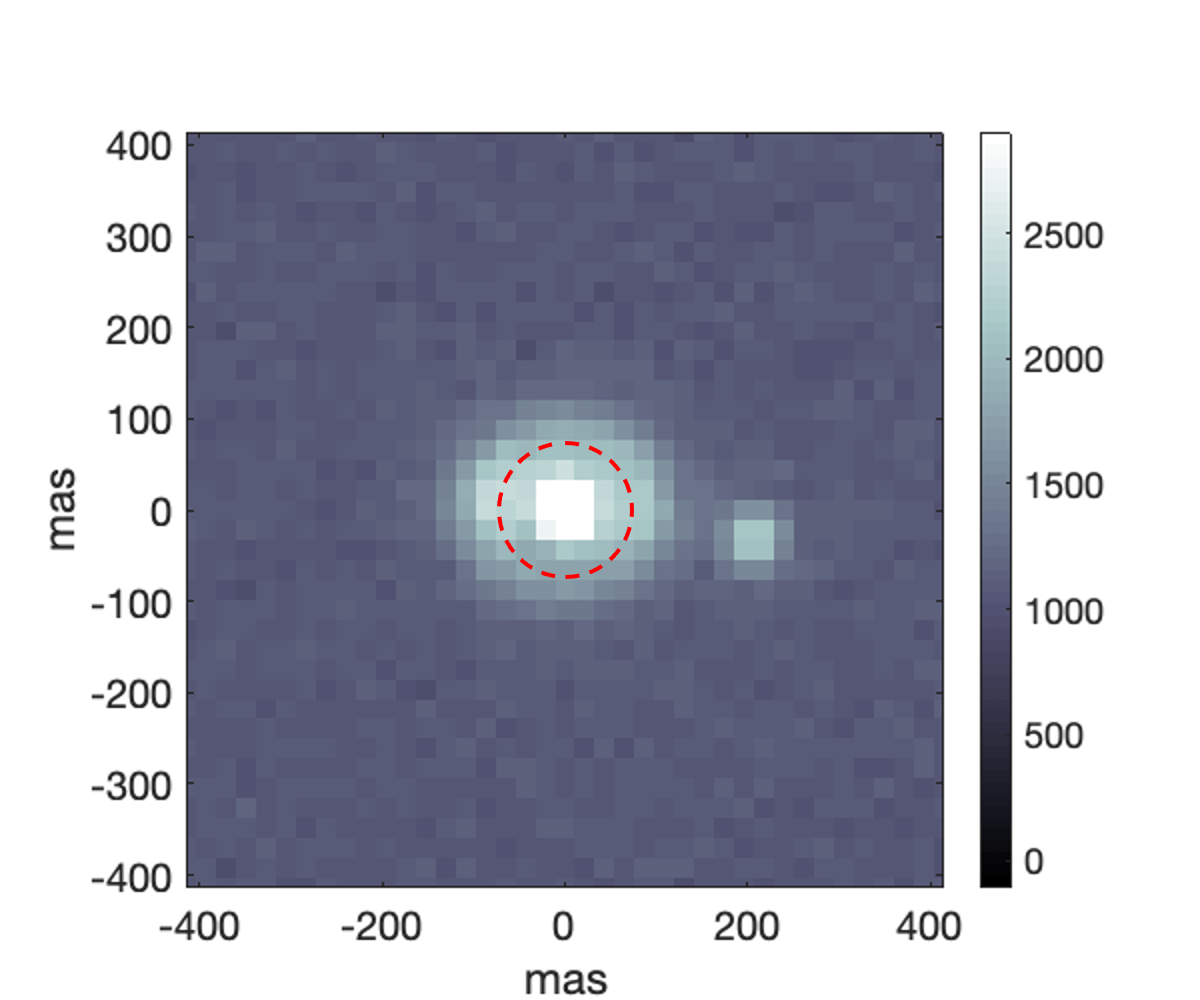} &
                    \includegraphics[trim=20 0 1 40,clip,width=0.48\textwidth]{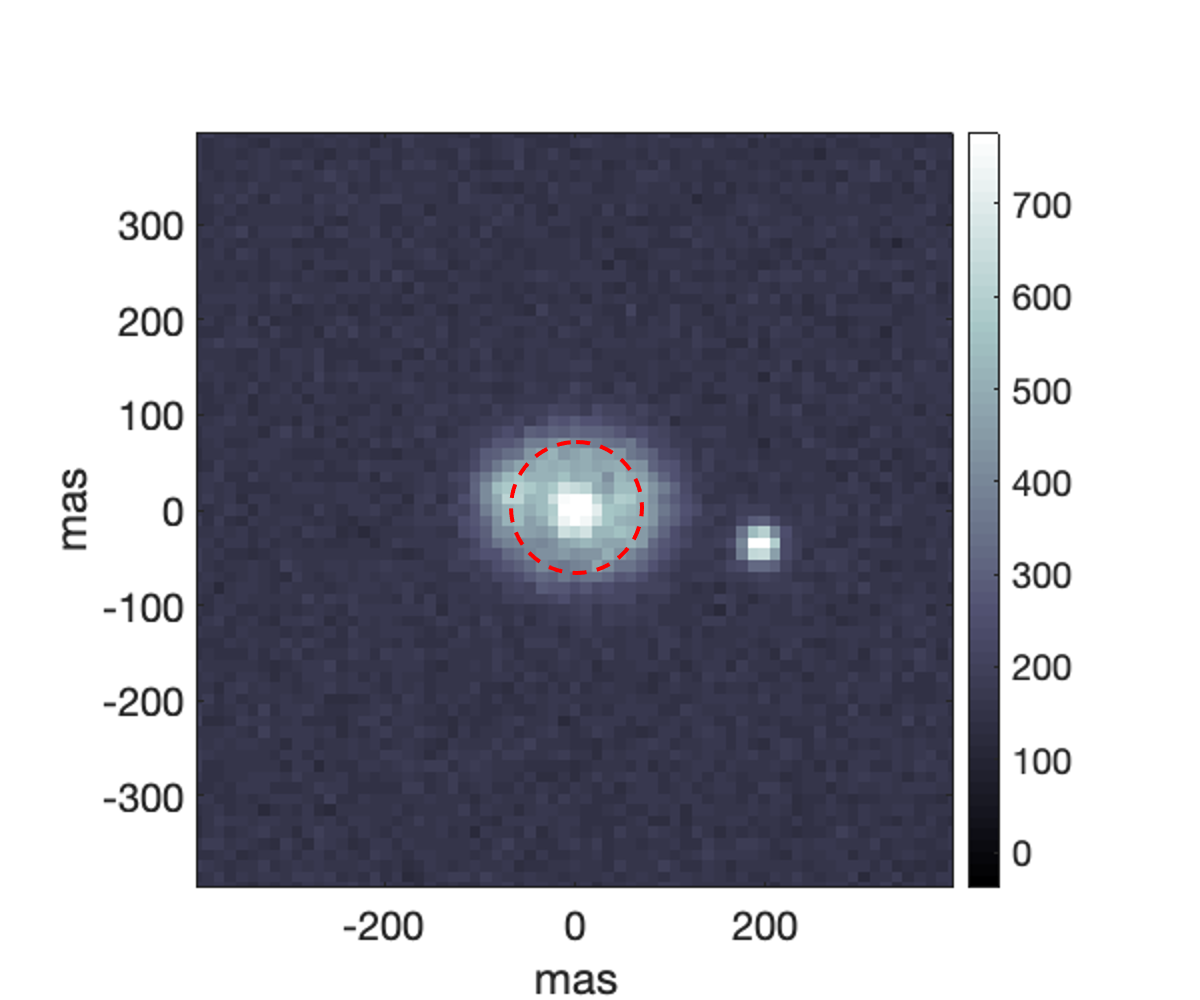} \\
                    \includegraphics[trim=20 0 1 40,clip,width=0.48\textwidth]{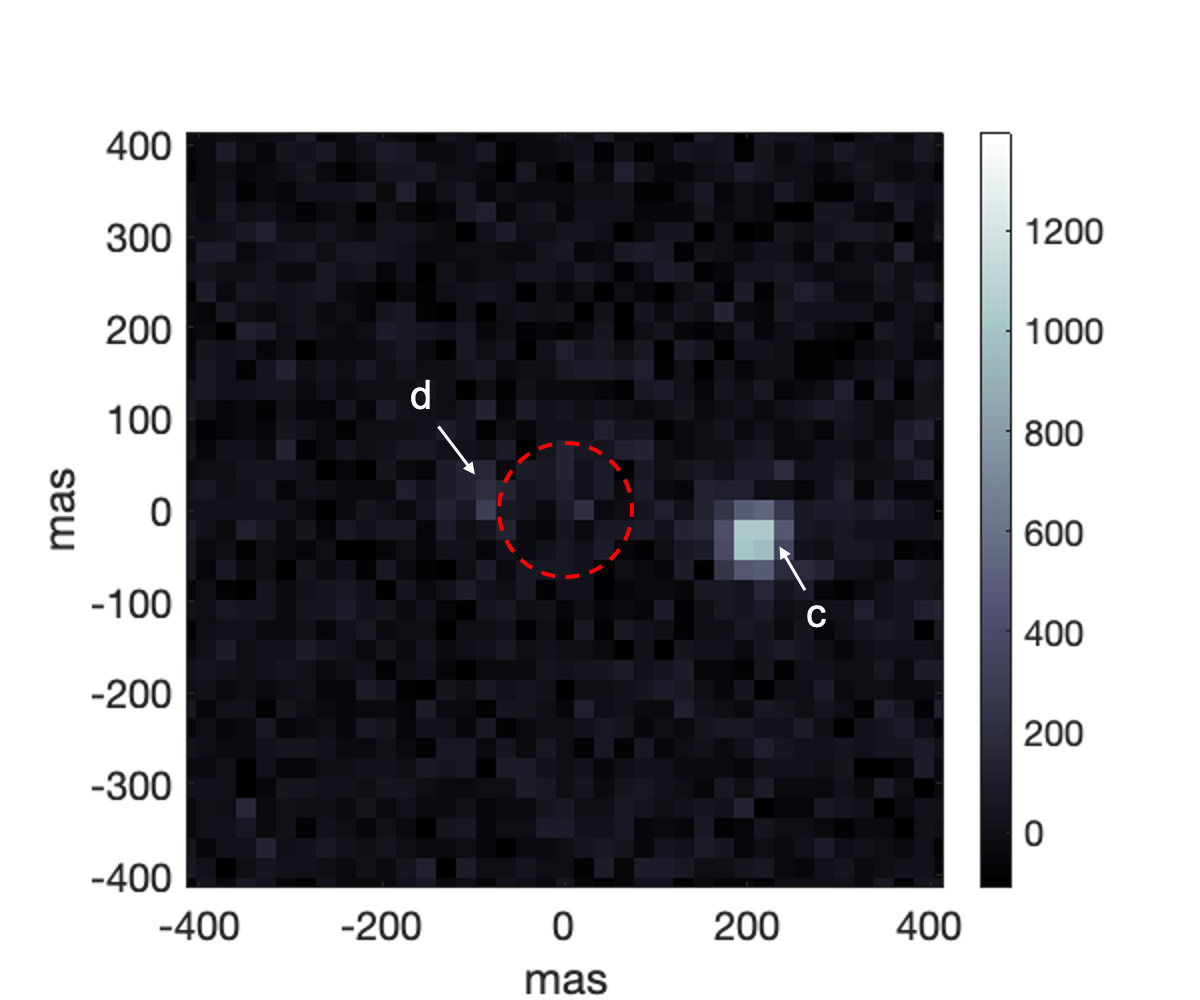} &
                    \includegraphics[trim=20 0 1 40,clip,width=0.48\textwidth]{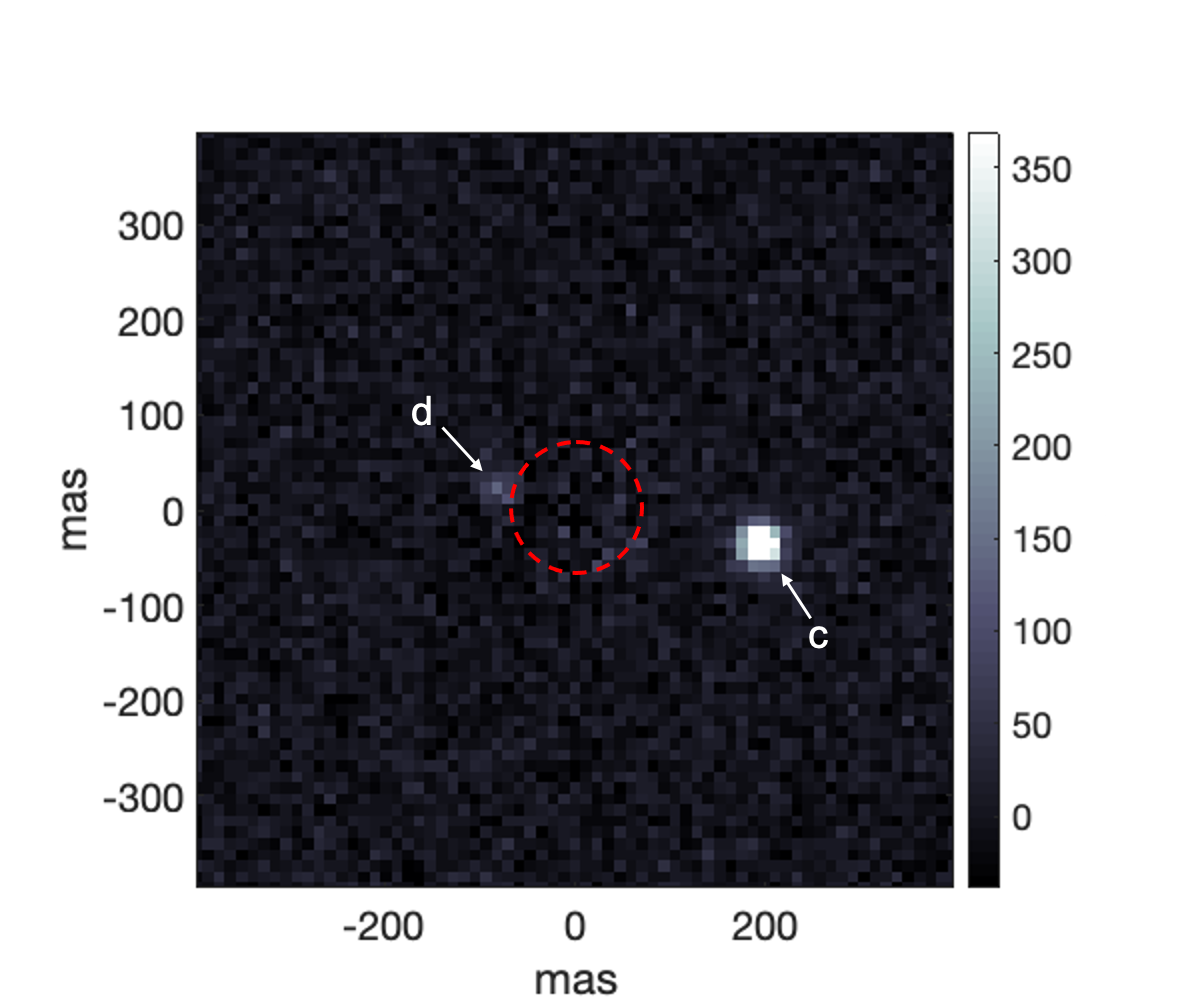} \\
                    \includegraphics[trim=11 0 5 30,clip,width=0.48\textwidth]{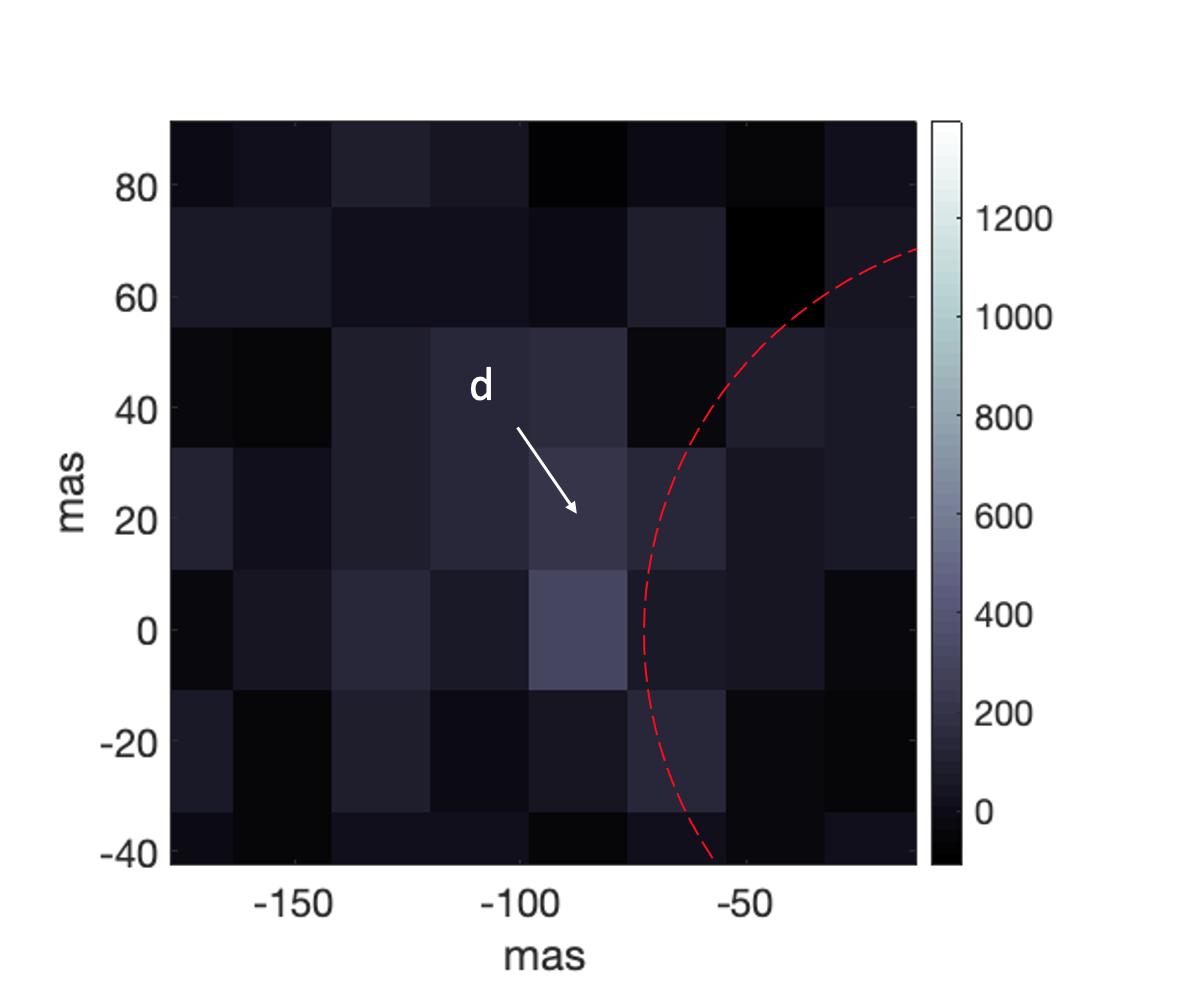} &
                    \includegraphics[trim=10 0 1 30,clip,width=0.48\textwidth]{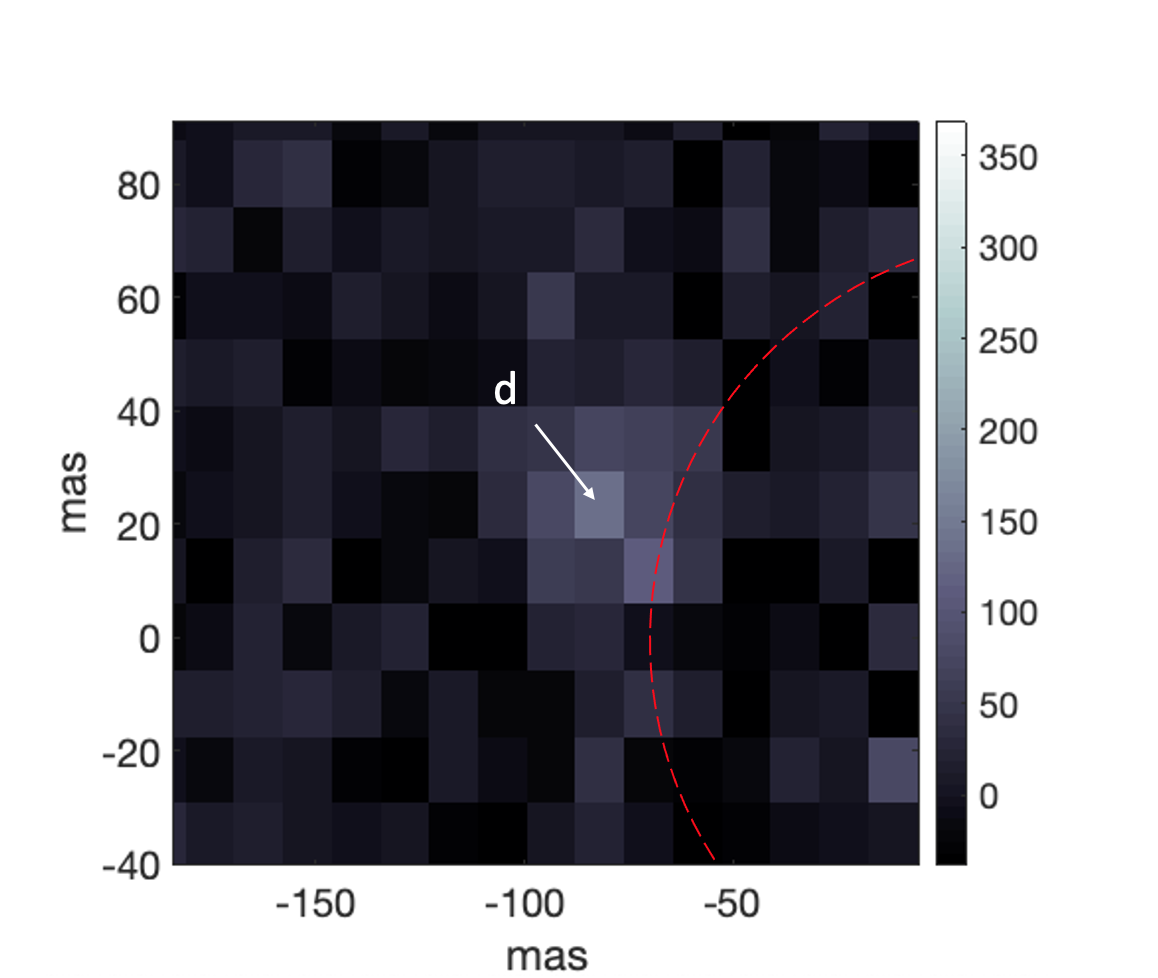} \\
                \end{tabular}
            \end{center}
            \caption{Left column images show the SISTER simulation for SRP of the planetary scenario around HD 134987 with detector's pixel scales of 21.8 mas/pix. Right column images show similar simulation for HabEx with pixel scales of 11.7 mas/pix. The integration time was chosen to recover a SNR = 5 for the fictitious planet. Top panel images are simulation with all the components both astrophysical and instrumental. Middle panel images show only the planetary signals after the subtraction of unbiased non-planetary signals, see text. Lower panel images show the zoomed in view around the location of planet d, which is the fictitious planet we injected into the system. The colorbars representing the level of brightness are in units of integrated photoelectrons. Arrow heads in the lower panel point to the exact position of the fictitious planet in those images. Our fictitious planet is not clearly visible before the noise subtraction in the upper panel, but is revealed afterwards. Planet b is not visible in all cases since it is inside the IWA$_{0.5}$ of the instrument. Planet c is clearly visible in all situations.}
            \label{fig:sister}
        \end{figure*}

        The simulations assumed instrumental parameters consistent with current best estimates of starshade non-ideal performance \citep{Seager2019,Gaudi2020,Hildebrandt2021,Romero-Wolf2021}.
        We included imperfect starshade petal edges from manufacturing and deployment errors consistent with mission requirements leaving a residual starlight equivalent to a 
        $1.4 \times 10^{-10}$ contrast ratio, a few times higher than the ideal limit of SRP and HabEx. Solar glint~\citep{McKeithen2021} was computed under the assumption that the starshade edges are coated with an anti-reflection coating similar to one that has been tested in the 
        laboratory. The Sun is 60$^{\rm o}$ from the starshade normal, which is a median value among planned observations, and is oriented at 45$^{\rm o}$ from horizontal. We included pointing jitter of the telescopes (14 mas for SRP and 2 mas for HabEx) consistent with expected in-flight performance of the Attitude Control Systems. Regarding the optical performance of SRP and HaBEx, we incorporated the optical throughput and expected loses as described in the corresponding reports. The performance of \textit{Roman} detectors were assumed to be in its end-of-life (EOL) condition since the SRP is not expected to be paired up with \textit{Roman} until near the end of the telescope mission. HabEx instrumental conditions were assumed to be beginning-of-life (BOL). In both cases, we assumed a EMCCD detector with an instrumental setup and single exposure consistent with each mission specifications, including quantum efficiency, dark current, clock induced charge noise, read-out noise and other loss effects. The simulations also included local zodiacal light (with a surface brightness of 23 V mag/arcsec$^2$), as well as a smooth exozodiacal cloud resulting from a simulation performed with ZODIPIC \citep{Kuchner2012}\footnote{We used a version of ZODIPIC with a more accurate handling of the forward scattering term, see \citep{Roberge2017}}, a zodiacal cloud simulation tool that incorporates observational data from the solar system. The dust density was chosen to be 5 times that of the solar system and we chose a similar asymmetry factor as in the solar system to simulate the forward scattering effect, with $g=0.2$, where g is the asymmetry factor for the Henyey-Greenstein phase function, see e.g. section 3.2 in~\citep{Stark2011}. For the background object, we did not add any galactic background star but we selected the brightest background galaxy found in the Haystack data cubes \citep{Roberge2017}. The resulting starshade images of both SRP and HabEx with all components are shown in the upper panel of Figure~\ref{fig:sister}.
        
        The total integration time was chosen as to provide a signal-to-noise ratio (SNR) of 5 for both SRP and HabEx to show what a detection would look like, shown in the middle panel of 
        Figure~\ref{fig:sister} as well as in the lower panel where the images zoomed in around the 
        location of the fictitious planet. The total integration time for SRP is of the order of 7 days, whereas for HabEx is of the order of 6 hours. The main factors that reduce the integration time between HabEx and SRP are by relevance: (i) the
        optical collecting area in HabEx, which is about 3.5 times larger than in SRP; (ii) the broader bandwidth in HabEx's visible channel (450-975 nm) compared to the SRP's 425-552 nm channel, although the actual value depends on the star's spectrum and planet's albedo;
        (iii) the optical efficiency of the system: HabEx approximately doubles that of SRP;
        (iv) better detector properties (quantum efficiency and noise properties); and (v) a 
        better disambiguation of the signal of the exoplanet from that of 
        diffuse background sources in the case of HabEx, in particular, exo-zodiacal light, 
        due to HabEx's finer angular resolution. Summarizing all these factors in the case of the fictitious planet, the photon detection rate in HabEx's visible channel is about 9 times that of SRP's 425-552 nm channel.
        The calculation of the SNR was performed using 
        aperture photometry under the assumption that an unbiased, yet noisy, 
        estimation of all non-planetary signal can be subtracted from the noisy 
        simulation with all components present. These non-planetary components are by
        relevance: the exozodiacal cloud, the local zodiacal light, 
        the solar glint~\citep{McKeithen2021} and residual scattered starlight 
        from the imperfect starshade. For instance, when the fictitious planet is near the IWA with
        a planet-star flux ratio of $\sim 2.5\times10^{-10}$, see Figure~\ref{fig:visibility}, 
        the relative weight of each component compared to the total signal in a box with side
        length 1 FWHM of the average PSF response is, approximately: 
        exozodiacal cloud ($\sim60\%$), local zodiacal light ($\sim15\%$),
        solar glint and residual starlight from the imperfect starshade 
        ($\sim18\%$ in the case of SRP,
        and $\sim9\%$ in the case of HabEx), and the planetary signal 
        ($\sim7\%$ in the case of SRP and $\sim16\%$ in the case of HabEx). 
        Notice that the spatial morphology of each component 
        is different. In particular, the brightness of 
        the local zodiacal light is the same across the image given the
        small angular size of the simulation, except for intrinsic fluctuations, i.e., its shot noise. The imaging simulation software SISTER~\citep{Hildebrandt2021} takes care of each of them adequately. In practice, this is equivalent to a simulation with the planetary signal and its shot noise component, together with the shot noise contribution from the non-planetary signal and the detector noise, both increased by a factor of $\sim \sqrt{2}$ because all noise sources are assumed to 
        follow either a Poisson or Gaussian probability distribution (Figure~\ref{fig:sister}). This is consistent with the methodology in SRP and HabEx reports, as well as with other similar investigations \citep{Robinson2016,Feng2018,Seager2019,Damiano2020,Gaudi2020,Romero-Wolf2021}. Our estimates are consistent with the ones derived in such publications. Since the reference flux ratio value for the fictitious planet is greater than or equal to $2.25\times10^{-10}$, which is higher than the residual starlight fluctuations due to the imperfect starshade, $\sim 1.4\times 10^{-10}$, the SNR could be increased at the expense of increasing the integration times. Clearly, that is not an issue for HabEx.
    
%%%%%%%%%%%%%%%%%%%%%%%%%%%%%%%%%%%%%%%%%%%%%%%%%%%%%%%%%%%%%%%%%%%%

\section{Discussion}
\label{discuss}

Although SRP and HabEx both have similar required contrast ratio limit of 1$\times$10$^{-10}$, the imaging simulation results show that HabEx is clearly superior. At similar flux ratio levels, the integration time required to successfully retrieve the signal of the fictitious planet with a SNR of 5 is much shorter for HabEx. As explained in Section \ref{DIretrieve}, HabEx's larger, unobstructed telescope, its broader passband in the visible channel and other instrumental advantages compared to SRP result 
in a higher integrated flux of about 9 times that of SRP in the case of the fictitious planet around HD134987. With an integration time of around 7 days for SRP, imaging of our fictitious sub-Neptune planet proves to be a challenging task at a distance of 26.2 pc. A smaller terrestrial planet at similar distance and separation from the host star would result in even lower planet-star flux ratios, making the required integration time for planetary detection much longer, possibly on the order of 30 days or more, which is not feasible for direct imaging observations. Therefore, imaging an Earth like counterpart at this distance with SRP is clearly unfavorable and the detection and characterization of small terrestrial planets are likely to be limited to within 20 pc or closer, until the launch of HabEx. 

Several assumptions were made and some of them may affect the result of this work. We discuss the implications of changing those assumptions here:
\begin{itemize}
    \item Since the inclinations of the two known planets in the HD 134987 system are unknown, here we assumed an inclination of 45$^{\circ}$ for all three planets, including our fictitious planet. Varying this value would have consequences on the flux ratios on all planets. If the true inclination of the system is less than 45$^{\circ}$ making the orbits more face-on, the fictitious planet would have a larger fraction of its orbit bright enough for detection, even though the peak brightness would not be as bright as the 45$^{\circ}$ case. However, too small of an inclination angle would likely eliminate the likelihood of the presence of such a planet in this system since larger planetary masses boosted by the $\sin i$ factor would diminish the already narrow dynamically allowed parameter space (Figure \ref{fig:megno}). On the other hand, if the true inclination is more towards edge-on, more parts of the orbit would be within the IWA or IWA$_{0.5}$, costing the chance of detection in this case, which is not ideal for direct imaging. In addition, if a different inclination value was selected for the study, or the true inclination of the system obtained through other methods in the future indicates a drastically different angle, the size of ``The Region" in Figure \ref{fig:threshold} would vary accordingly as well as the allowed paramter space for the fictitious planet. The RV curves would shift further towards smaller or larger separation for more face-on or edge-on orbits, respectively. In the case of HD 134987, ``The Region" exists in all cases and our fictitious planet is valid for the purpose of this study, which is to probe the parameter space in between RV and imaging sensitivities, no matter what inclination was picked. However, this might not be true if similar studies were to be carried out for other systems and a more careful selection of the inclination angle might be needed in other cases.
    \item The physical and orbital parameters of the fictitious planet make an impact on its detectability for direct imaging. Here we assumed a near zero eccentricity for our planet for convenience. However, as pointed out in \citet{Kane2013}, at the same separation for the star, high eccentricity planets could improve its chance of detection. In addition, the fictitious planet could take on different radius values and thus different mass values. The allowed radius range for the purpose of this work according to Figure \ref{fig:threshold} is roughly between 1 and 2.5 Earth radii, for which the masses are predicted to be between 1 and 7 Earth masses. Higher radius for the planet would have increased reflected flux, making them easier for detection, but the location of the planet has to be farther away from the star in order to stay in between the RV and direct imaging detection thresholds (Figure \ref{fig:threshold}) that we are interested in. The result of the two might not necessarily make the fictitious planet any more favorable for detection than the current configuration. When working on known planets that only have mass information, either in the HD 134987 or other systems, the deviation from the true value using any mass-radius relationship may have an impact on the detectability if such a planet lies near the detection threshold of any imaging instruments.
    \item The simulation of realistic geometric albedo of the two known planets and the fictitious planet is beyond the scope of this paper. Instead, we assumed geometric albedo of 0.5 for the two know giant planets and 0.3 for the fictitious planet. This assumption is based on previous works such as \citet{Cahoy2010} and \citet{Kane2010} on estimating the albedo spectra of solar system analogs and albedo variation of giant planets with respect to orbital distances. More careful albedo treatment can be obtained through atmospheric modeling for each planet. Higher geometric albedo values for planets would result in higher chance of detection.
\end{itemize}

It is worth noting that this work is by no means dictating the presence of additional yet to be discovered planet in the HD 134987 system. Rather, it shows that such discovery is entirely possible with future direct imaging missions thanks to the complementary detection sensitivity of direct imaging to RV's. Although the work presented here could only be applied to one system at a time, in this case the HD 134987 system, similar procedure could be carried out for other direct imaging systems with different parameters and assumptions. This in-depth look at a particular system with the combination of different detection sensitivites curves, along with realistic RV, imaging, and dynamical simulations, allows the parameter space within the system to be fully explored and provides unique system information that other statistical studies cannot offer. This is particularly the case when there are more than one known planets exist within the system, where the known bodies can provide constraints on the physical and orbital properties of the possible planetary candidate in between the orbits of known planets through dynamical simulations, or completely rule out such likelihood, therefore saving telescope time from searching for new exoplanet candidate from ruled out systems. Results from similar studies for other direct imaging systems, if carried out, could be valuable tools for determining systems with likelihood of hosting additional planets, predicting the best time and instruments to recover such potential planets, informing the target selection process of future imaging missions as well as maximizing the limited resources of starshade operation in the hunt for new planets.

\section{Conclusion}
\label{conclude}
In this work, we demonstrated the steps in trying to predict whether new planets that are below the RV detection thresholds can be discovered by future direct imaging missions in one particular case of the HD 134987 system, one of the farthest systems to be imaged by future missions. We used HD 134987 system as a test case and developed a detection threshold figure that combines sensitivity curves from both RV and direct imaging. This figure (Figure \ref{fig:threshold}) was constructed with the most optimistic scenario and does not take into account dynamical influence of known bodies in the system as well as the additional noise sources such as stellar jitter, starshade imperfection, local and exozodiacal light. Thus, a careful examination of the system is necessary to 
verify the detection limit. We have considered some detailed simulations that take into account system stability as well as realistic starshade images with an imperfect starshade within mission requirements. Based on Figure \ref{fig:threshold}, we were able to locate the parameter space to search for the location of the potential new planet through dynamical analysis. We assigned a sub-Neptune fictitious planet and injected it into the HD 134987 system for retrieval tests. RV periodogram search and modeling on a synthetic RV dataset returned negative results as expected, while direct imaging retrieval was able to successfully recover the signal of the fictitious planet. Through imaging simulations, we were able to directly compare the imaging capabilities of SRP and HabEx and the results show that imaging with HabEx is clearly superior and SRP is probably not capable of imaging low mass planets at distance around 26 pc. Since the majority of the future imaging candidates are RV planets, the synergy between RV and imaging becomes ever-increasingly crucial. These predicted directly imaged planets would provide potential discoveries of cold giant, sub-giant planets at long orbital periods, and even terrestrial planets in the HZ, increasing the science return of the missions, as well as opportunities for atmospheric characterization in the low incident flux regime.

%%%%%%%%%%%%%%%%%%%%%%%%%%%%%%%%%%%%%%%%%%%%%%%%%%%%%%%%%%%%%%%%%%%%

\section*{Acknowledgements}

The authors would like to thank Chris Stark for sharing an improved version of ZODIPIC, which was used to generate the input exozodiacal dust in section~\ref{DIretrieve}. The authors wish to also thank the referee 
for providing a swift and valuable response that greatly improved the presentation of this work. This research has made use the NASA Exoplanet Archive, which is operated by the California Institute of Technology, under contract with the National Aeronautics and Space Administration under the Exoplanet Exploration Program. Dynamical simulations in this paper made use of the REBOUND code which is freely available at \url{http://github.com/hannorein/rebound}. 

%%%%%%%%%%%%%%%%%%%%%%%%%%%%%%%%%%%%%%%%%%%%%%%%%%%%%%%%%%%%%%%%%%%%

\software{RadVel} \citep{Fulton2018}, 
{RVSearch} (Rosenthal et al. submitted)
{REBOUND} \citep{Rein2012},
{SISTER} \url{sister.caltech.edu}~\citep{Hildebrandt2021}

%%%%%%%%%%%%%%%%%%%%%%%%%%%%%%%%%%%%%%%%%%%%%%%%%%%%%%%%%%%%%%%%%%%%
%\bibliographystyle{aasjournal}
%\bibliography{references}

%%%%%%%%%%%%%%%%%%%%%%%%%%%%%%%%%%%%%%%%%%%%%%%%%%%%%%%%%%%%%%%%%%%%

\end{document}